\documentstyle[twocolumn,aps,epsfig]{revtex}
\begin{document}
\draft
\title{Surface Phase Diagrams for Wetting on Heterogenous Substrates}

\author{C.\ Rasc\'{o}n$^{\star}$ and A.\ O.\ Parry}
\address{Department of Mathematics, Imperial College\\
180 Queen's Gate, London SW7 2BZ, United Kingdom}
\maketitle

\begin{abstract}
We propose a simplified description of fluid adsorption on heterogenenous 
micropatterned substrates. Using this approach, we are able to rederive 
results obtained earlier using effective interfacial Hamiltonian methods 
and predict a number of new examples of surface phase behaviour for both
singly and periodically striped substrates. In particular, we show that, 
for a singly striped system, the manner in which the locus of surface 
unbending phase transitions approaches the pre-wetting line of the 
infinite pure system, in the limit of large stripe widths, is non-trivial 
and sensitive to several characteristic lengthscales and competing 
free-energies. For periodic substrates, we investigate finite-size 
deviations from Cassie's law for the wetting temperature of the 
heterogeneous system when the domain sizes are mesoscopic.

\end{abstract}

\section{Introduction}

Fluid interfacial phenomena on
flat, homogeneous substrates have drawn enormous attention
over the last few decades \cite{Review}. However,
present experimental methods allow an ever increasing control over 
the shape and chemical composition of solid surfaces and pose more
demanding challenges for both theory and experiments. These advances
make the study of adsorption on non-planar and heterogeneous
substrates inevitable. Recent experimental and theoretical
studies have shown that structured substrates exhibit a variety of
novel adsorption properties which not only promise to be of importance 
to future technologies such as microfluidics \cite{Microfluidics},
but are also of fundamental interest to statistical physics \cite{Geometry}.
In particular, the effect of chemical inhomogeneities has been
addressed recently in different contexts, for example: contact angles of
liquid drops \cite{Drelich,Swain,Wood}, droplet spreading \cite{Adao},
morphological phase transitions \cite{Lenz}, three phase contact line \cite{BD1},
Cassie's law \cite{Urban,Henderson},
drop shapes \cite{PMR}, construction of magnetic materials \cite{Palacin}, 
microscopic packing \cite{Frink}, liquid channels \cite{Gau},
polymer blends \cite{Polymer1,Polymer2}, and
dewetting \cite{Kargupta}, among others.

In this paper, we concentrate on the equilibrium wetting 
properties of flat but chemically structured substrates.
Specifically, we study liquid adsorption in two representative
micropatterned systems: a homogeneous substrate with a single distinct 
chemical stripe
and a substrate comprising parallel stripes of two different materials.
These systems have been partially studied by other authors
with the help of density functional theory \cite{Frink,BD2}
and effective interfacial methods \cite{BDP,BD3,Granada}.
Nonetheless, the intrinsic complexity of those formulations does
not allow a thorough exploration of the vast space of
possible surface phase behaviour these systems offer.
We propose here a minimal model to understand the general
wetting properties of these systems. The model, described
in section II, appears to
capture the essential physics of the problem in a qualitative
manner and recovers former results obtained with more 
sophisticated methods for particular cases. 
The great advantage of the minimal model is that it allows us
to obtain (sections of) global surface phase diagrams and
explore what kind of surface phase behaviour is possible in 
such heterogeneous systems. As we shall show, in spite of
the simplicity of the model, the ensuing surface phase behaviour
is extremely rich.
In section III, the model is applied to a substrate with a
single stripe of a different material, while in
section IV, a periodic substrate made of parallel stripes
of two materials is studied. We finish our article with a
summary of our main conclusion and a discussion of further work.

\section{Minimal Model}
Consider two flat substrates made of different materials
(labelled 1 and 2) whose adsorption properties can be described, when
homogeneous,
by effective potentials $W_1(\ell)$ and $W_2(\ell)$, respectively
\cite{Weff}. When each of these substrates is exposed to a near
saturated gas, a liquid layer is adsorbed on its surface, whose thickness
$\ell_1^\pi$ (or $\ell_2^\pi$) is given by the position of
the absolute minimum of the corresponding effective potential
(the superscript $\pi$ denotes behaviour
characteristic of the pure homogeneous substrate). 
We suppose the wetting temperatures of each of these systems
are  $T_{\mathrm{w}}^{(1)}$ and $T_{\mathrm{w}}^{(2)}$, respectively.
Consider now a flat substrate $\Lambda$
consisting of domains $\Lambda_1$ and $\Lambda_2$ (not necessarily connected)
made of materials 1 and 2, respectively.
%($\Lambda_1\cup\Lambda_2=\Lambda$, $\Lambda_1\cap\Lambda_2=\emptyset$).
When this microprinted substrate is exposed to a gas,
a liquid layer is adsorbed on its surface, whose equilibrium thickness 
profile $\ell(x,y)$ must show considerable spatial dependence 
(especially strong close to the boundaries between the domains).
For mesoscopic domains, this dependence can be captured by 
an effective interfacial description:
the equilibrium profile of the adsorbed liquid layer will minimise the
following free-energy functional

\begin{eqnarray}
\label{zero}
{\cal H}[\ell]=\int_{\Lambda}\!\!d{\bf r}\,\;
{\Sigma\over 2}\;\left(\nabla\ell\right)^{\,2}+
\int_{\Lambda_1}\!\!\!d{\bf r}\,\;W_{1}(\ell)+
\int_{\Lambda_2}\!\!\!d{\bf r}\,\;W_{2}(\ell),
\end{eqnarray}
where $\Sigma$ is the liquid-vapour interfacial tension.
This model assumes a sharp crossover in the local form
of the effective potential at the domains boundaries
and is itself an approximation to a more realistic
description in which this crossover is smooth \cite{BDP}.
However, this assumption certainly does not alter the global
structure of the surface phase behaviour \cite{BD1}.
To continue, and for the sake of simplicity, we only
consider systems translationally invariant along the
$y$ axis (see Fig.\ 1). Therefore, the equilibrium profile will
follow from the minimisation of the somewhat simpler functional
\begin{equation}
\label{two}
H[\ell]=\int_{\cal L}\!\!dx\,\;
{\Sigma\over 2}\;\dot{\ell}^{\,2}+
\int_{{\cal L}_1}\!\!\!dx\,\;W_{1}(\ell)+
\int_{{\cal L}_2}\!\!\!dx\,\;W_{2}(\ell),
\end{equation}
where the dot represents differentiation with respect to $x$,
and $\cal L$, ${\cal L}_1$ and ${\cal L}_2$ are the appropriate linear domains
corresponding to sections of $\Lambda$, $\Lambda_1$ and $\Lambda_2$, respectively.
This approach has been employed by a number of authors to obtain
numerical solutions of the equilibrium profile for a substrate with
a single stripe of a different material
(Fig.\ 1(a))\cite{BDP,BD3,Granada}.
Their results show that if the width of the stripe $L_1$ is very large,
the thickness of the adsorbed layer is essentially the same as in
the pure systems within each of the domains ($\ell_1^\pi$ in $\Lambda_1$,
$\ell_2^\pi$ in $\Lambda_2$), and varies abruptly
in the domain borders. In other words, for large stripe widths, 
the minimisation
of (\ref{two}) is achieved by a local minimisation of the second
and third terms of the functional. The (positive) contribution of
the first term in (\ref{two}), which originates in the domain borders,
is practically negligible. We can estimate this latter contribution by
noticing that the characteristic length for lateral variations of
the interface profile is given roughly by the transverse correlation lengths
$\xi_{\parallel}^{(i)}\!=\!\sqrt{\Sigma/W''(\ell_{i}^\pi)}$  of the 
homogeneous substrates $i\!=\!1,2$. Thus, for a single step,
we can write
\begin{equation}
\label{three}
\int\!\!dx\,\;{\Sigma\over 2}\;\dot{\ell}^{\,2}
\;\approx\;
%(\xi_{\parallel}^{(1)}\!+\!\xi_{\parallel}^{(2)})\;
%{\Sigma\over 2}\left(\frac{\ell^{\pi}_1\!-\!\ell^{\pi}_2}
%{\,\xi_{\parallel}^{(1)}\!+\!\xi_{\parallel}^{(2)}}\right)^{\,2}\!\!\!\!=
{\kappa\over 2}\,\left(\ell^{\pi}_1\!-\!\ell^{\pi}_2\right)^{\,2}
\end{equation}
with $\kappa\equiv\Sigma/(\xi_{\parallel}^{(1)}\!+\!\xi_{\parallel}^{(2)})$.
For smaller values of $L_1$, however,
this energetic contribution becomes comparable
to that arising from the second term of (\ref{two}) and the
interface adopts distinct configurations in order to minimise the
free-energy of the system. This gives rise to certain surface phase transitions,
denoted surface condensation or {\it unbending} transitions,
in which different liquid interfacial configurations
coexist\cite{BDP,BD3,Granada,RPS}. In order to calculate the loci of
these transitions in the $T$-$\mu$ diagram, the free-energy functional
(\ref{two}) must be minimised for different values of the temperature $T$
and the chemical potential $\mu$. Although this has been done for a
specific case\cite{BD3}, the vast space of parameters involved and the
arduous task of minimising the functional (\ref{two}) does
not allow a thorough exploration of the physics of the system and
impedes a systematic study of more complex systems, for instance,
a periodic array of stripes (see Fig.\ 1(b)).

Here we propose a minimal model to describe the overall 
phenomenology of the above mentioned surface {\it unbending} transitions
in a simple and illuminating manner.
To do that, we approximate the free-energy {\it functional} (\ref{two})
by a free-energy {\it function} of a finite number of variables.
By doing that, we lose the detailed description of the shape of the 
adsorbed layer (which follows from the minimisation of Eq.\ (\ref{two})),
but, as explained here, we capture the fundamental physical mechanism 
for the surface phase transitions underlying (\ref{two}).
To start, we define a collective coordinate $\ell_i$ for every connected 
region of each of the domains. This coordinates will account for the 
{\it average} thickness (in a loose manner)
of the adsorbed layer within that connected region
and it is defined implicitly as follows
\begin{equation}
\label{four}
L_i\,W_\alpha(\ell_i)=
\int_{L_i}\!\!\!dx\,\;W_\alpha(\ell),
\end{equation}
$L_i$ being the width of the 
connected region and $\alpha\!=\!1$ or 2,
depending on the domain that region belongs to.
Within the minimal model,
we consider these collective coordinates as independent
entities, to be varied in order to minimise the
free-energy. This accounts for the second and third integrals of 
(\ref{two}) which are replaced by a sum over terms 
$L_i\,W_\alpha(\ell_i)$, one for each connected region.
To obtain a full description
in terms of these variables, we simply estimate the contribution 
of the first term of (\ref{two}) similar to that shown in
Eq.\ (\ref{three}) but using the collective 
coordinates. Thus, the free-energy functional is reduced 
to a function of a small number of real variables.
Specifically, for the periodic array of stripes (Fig.\ 1(b)),
we write the free-energy per period as
\begin{equation}
\label{five}
H(\ell_1,\ell_2)=\kappa\left(\ell_1\!-\!\ell_2\right)^{\,2}+
L_1\,W_1(\ell_1) + L_2\,W_2(\ell_2).
\end{equation}
In accordance with our aim of developing a minimal description, 
we consider $\kappa$ constant throughout the paper (independent 
of the temperature and the chemical potential). 

The free-energy for a substrate with a single 
stripe (Fig.\ 1(a)) can be obtained from
(\ref{five}) in the limit $L_2\!\rightarrow\!\infty$. In this
case, the last term of (\ref{five}) becomes the most important
and it must be minimised fully, yielding $\ell_2\!=\!\ell_2^\pi$.
Substracting this (infinite) constant contribution from the free-energy,
the average thickness of the
adsorbed layer on the stripe will follow from the minimisation of 
\begin{equation}
\label{six}
H(\ell_1)=\kappa\left(\ell_1\!-\!\ell_2^\pi\right)^{\,2}+
L_1\,W_1(\ell_1).
\end{equation}

The generalization of this procedure to more sofisticated geometries
is straightforward. As an example, the average
thickness $\ell_1$ of the adsorbed layer
on a circle of radius $R$ made of material 1 on a substrate made of
material 2 could be obtained from the minimization of
\begin{equation}
\label{seven}
H(\ell_1)=\pi R\,\kappa\left(\ell_1\!-\!\ell_2^\pi\right)^{\,2}+
\pi R^2\,W_1(\ell_1),
\end{equation}
where an appropriate $\kappa$ now accounts for the 
energy due to the curvature of the interface.
It is likely that this procedure will fail to work for certain
intricate shape domains where more complicated behaviour may be
expected.

We stress here that this approach does not intend to be a
quantitative description of these phenomena
but to capture the essential features of heterogeneous
adsorption, which, due to the complexity of the available models,
remain mainly unexplored.

\section{Single Stripe}

\subsection{General Considerations}
\label{sec1}

Here we apply the model described above to a homogeneous flat 
substrate with an infinite length stripe of width $L_1$ (Fig.\ 1(a)).
Before presenting the results, however,
we make a number of pertinent remarks concerning
general aspects of the surface phase behaviour:

\begin{itemize}

\item For small values of the stripe width,
corresponding to the limit $L_1\!\rightarrow\!0$,
we expect that the thickness of the layer adsorbed on the stripe will 
tend to the thickness of the liquid layer absorbed on the
outer substrate, {\it i.e.} $\ell_1\!\rightarrow\!\ell_2^\pi$.

\item If the stripe width is large, corresponding to
the limit $L_1\!\rightarrow\!\infty$,
the thickness of the layer adsorbed on the stripe will tend to the 
thickness of the liquid layer absorbed on a homogeneous 
flat substrate of material 1, {\it i.e.} $\ell_1\!\rightarrow\!\ell_1^\pi$.

\item If the thickness of the adsorbed layers on both infinite
substrates is the same ($\ell_1^\pi\!=\!\ell_2^\pi$), the functional
(\ref{two}) is fully minimised by a flat solution
$\ell(x)\!=\!\ell_1^\pi\!=\!\ell_2^\pi$.
The loci of the points for which $\ell_1^\pi\!=\!\ell_2^\pi$ 
constitute a line in the $T\!-\!\mu$ phase diagram and separates
interfacial profiles with different convexity\cite{Flat}.
Observe that the flat solution corresponds to
the minimisation of {\it each} term of the functional and cannot be
improved by any other configuration. It follows that such a solution will be
the global equilibrium for arbitrary values of the stripe width $L_1$.
In turn, this implies that no first-order surface coexistence line
can cross the line defined implicitly by
$\ell_1^\pi\!=\!\ell_2^\pi$. As explained below, this simple observation
severely constrains the possible topology of the surface phase diagram.
Intriguingly, the loci of points satisfying $\ell_1^\pi\!=\!\ell_2^\pi$
represents surface phases which are fully homogeneous on the
heterogeneous substrate.

\item Any first-order phase boundary corresponding to a line
of coexistence in the $T\!-\!\mu$ plane satisfies 
a Clausius-Clapeyron-like equation \cite{Schick}:
\begin{equation}
\label{clacla}
\left(\frac{\mathrm{d}\mu\,}{\mathrm{d}T}\right)_{\mathrm{coex}}\!\!=
-\frac{\Delta S}{\Delta\Gamma},
\end{equation}
where $\Delta S$ and $\Delta\Gamma$ are the differences in
surface entropy and adsoption between the coexisting phases,
respectively, which follow in the usual manner from the
free-energy of the system $\omega(T,\mu)\!=\!\min H$ as
$S\!\equiv\!-(\partial\omega/\partial T)_{\mu}$ and
$\Gamma\!\equiv\!-(\partial\omega/\partial\mu)_{T}$.

For our purposes, it is more convenient to measure
the slope of this line with respect to that of the
liquid-vapour coexistence curve $\mu_{\scriptscriptstyle LV}(T)$.
Thus, we rewrite Eq.\ (\ref{clacla}) as
\begin{equation}
\label{clacla2}
\left(\frac{\mathrm{d}\,\Delta\mu\,}
{\mathrm{d}\,\Delta T}\right)_{\mathrm{coex}}\!\!=
-\frac{\Delta\widetilde{S}}{\Delta\widetilde{\Gamma}},
\end{equation}
where $\Delta\mu\!\equiv\!(\mu_{\scriptscriptstyle LV}\!-\!\mu)$,
$\Delta T\!\equiv\!T-T_{\mathrm{w}}$ (for a given wetting temperature),
$\widetilde{S}\!\equiv\!-(\partial\omega/\partial\,\Delta T)_{\Delta\mu}$
and $\widetilde{\Gamma}\!\equiv\!-
(\partial\omega/\partial\,\Delta\mu)_{\Delta T}\!=\!-\Gamma$.
\end{itemize}

\subsection{Minimal Model}

The equilibium value of the {\it average} thickness $\ell_1(T,\mu,L_1)$
follows from the minimisation of the minimal free-energy (\ref{six}),
yielding
\begin{equation}
\label{eight}
2\kappa\left(\ell_1\!-\!\ell_2^\pi\right)+ L_1\,W_1'(\ell_1)=0.
\end{equation}
This equation has a straightforward graphical interpretation
from which one can elegantly glean the possible surface phase
behaviour. To see this, we rewrite the equation as
\begin{equation}
\label{nine}
W_1'(\ell_1)=\frac{2\kappa}{L_1}\left(\ell_2^\pi\!-\!\ell_1\right),
\end{equation}
corresponding to the intersection of a straight line with a curve.
The general form of the graphical construction is illustrated in
Fig.\ 2 for a specific choice of effective potential.
Note the following features:
\begin{itemize}
\item The model always shows at least one solution. Multiple solutions
correspond to the presence of first-order surface phase transitions.
\item The straight line has a negative slope inversely proportional
to the stripe width and necessarily crosses the horizontal
axis at $\ell_2^\pi$.
\item The curve always intersects the horizontal axis at $\ell_1^\pi$
(but may exhibit other crossings) and tends to
$\Delta\mu\,\delta\rho$ as $\ell_1\!\rightarrow\!\infty$.
Here, $\delta\rho$ represents the difference in bulk
densities of the coexisting liquid and gas states.
\item For any given temperature and chemical potential,
only three scenarios are possible within the graphical
construction. These correspond to the straight line
intersecting the curve at one, two or three points.
In the first scenario [Fig.\ 2(a) or (b)], the intersection
constitutes the unique solution of Eq.\ (\ref{nine}).
In the second case [Fig.\ 2(c)], the straight line must be tangential to
the curve at a point (representing a spinodal) and crosses the curve
at a second point, which constitutes the stable solution. 
In the third case, the central intersection represents
an unstable solution whilst the other two are mechanically stable. Among them, 
the stable (metastable) solution will have the lower (higher)
free-energy. If both solutions have the same free-energy,
two different thicknesses of adsorbed layer coexist at the wall-fluid interface
corresponding to a first order phase boundary [Fig.\ 2(d)] \cite{First}.
\item In the limit $L_1\!\rightarrow\!0$, the straight line is vertical
and the solution tends to the thickness of the liquid layer adsorbed
on the surrounding substrate, $\ell_1\!\rightarrow\!\ell_2^\pi$, as 
expected (see previous subsection). Although the model is not intended
to be accurate for very narrow stripes, note that it recovers the
correct result in this limit.
\item In the limit $L_1\!\rightarrow\!\infty$, the straight line is 
horizontal and the solution recovers the homogeneous substrate,
$\ell_1\!\rightarrow\!\ell_1^\pi$ (see previous subsection).
\item If $\ell_1^\pi\!=\!\ell_2^\pi$, both lines cross at the
same point of the horizontal axis and there is only one solution
independent of the value $L_1$ (see previous subsection).
\item There is a critical point for a certain combination of
the parameters of the problem. This point corresponds to the
straight line intersecting the curve tangentially at the point
of zero curvature (and negative slope). See Fig.\ 2(b). Thus,
within this approach, a necessary condition for the existence 
of a critical point is the presence of a point in the effective 
potential for which the third derivative is zero (the 
second derivative at that point being negative). This condition
can be used to determine a bound for the critical point (see below).\\
Due to the simplicity of the model, the conditions for criticality
can be written explicitly as
\begin{equation}
\label{tena}
W_1''(\ell_1)=-\frac{2\kappa}{L_1}
\end{equation}
and
\begin{equation}
\label{tenb}
W_1'''(\ell_1)=0,
\end{equation}
which, together with Eq.\ (\ref{nine}) determine the critical 
point completely. Observe that Eqs.\ (\ref{nine}) and (\ref{tena})
constitute together the conditions for the spinodals. 
\end{itemize}

\subsection{Second-order Wetting Stripe}
\label{sec2}

To begin we consider the case where the stripe corresponds to a material that
exhibits a continuous (critical) wetting transition in the limit of infinite
stripe size. No assumptions about the outer domain have to be made at this 
moment. Here we concentrate on the most experimentally relevant case of
long-ranged forces and write our model effective potential in
the usual manner \cite{Review}

\begin{equation}
\label{eleven}
W_1(\ell)= \frac{\Delta T}{\ell^{\,p}}+\frac{A}{\,\ell^{\,p+1}}+
\Delta\mu\,\delta\rho\;\ell,
\end{equation}
where we have rescaled the temperature such 
$\Delta T\equiv T-T_{\mathrm{w}}^{(1)}$, $A$ is a (positive)
Hamaker constant
and $p$ determines the range of the forces ($p\!=\!2$ for Van der Waals 
dispersion forces).

The effective potential (\ref{eleven}) models the continuous divergence of the
film thickness as the temperature is increased towards the wetting temperature 
$T_{\mathrm{w}}^{(1)}$ at $\Delta\mu\!=\!0$.
Observe that the third derivative of this potential consists of
only two terms and does not depend explicitly on the chemical potential.
Recalling the condition for criticality, Eq.\ (\ref{tenb}),
it follows that a critical point can only occur provided $\Delta T\!<\!0$.
In other words, if this system has a critical point, it
must occur necessarily at a temperature {\it below} the
wetting temperature $T_{\mathrm{w}}^{(1)}$.\\

The conditions for criticality follow immediately from Eqs.\ (\ref{nine})
to (\ref{tenb}):

\begin{equation}
\label{twelvea}
\ell_{1_{\scriptstyle C}}=\left[
\frac{A\,L_1\,(p+1)}{2\kappa}
\right]^{\frac{\scriptstyle 1}{\,\scriptstyle p+3\,}}
\end{equation}

\begin{equation}
\label{twelveb}
\Delta T_C = \frac{p+3}{p}\,\frac{A}{\,\ell_{1_{\scriptstyle C}}}
\end{equation}

\begin{equation}
\label{twelvec}
\Delta\mu_C\,\delta\rho=
\frac{2\kappa}{L_1}\left(\ell_2^\pi - \frac{p+3}{p+1}\,
\ell_{1_{\scriptstyle C}} \right)
\end{equation}

Figure 3 shows the phase diagram obtained for the effective
potential (\ref{eleven}) and different values of the stripe
width $L_1$. For the sake of simplicity, the
thickness of the adsorbed layer in the surrounding substrate,
$\ell_2^\pi$, has been considered constant (independent of
the temperature and the chemical potential). This approximation
represents a system whose wetting temperature $T_{\mathrm{w}}^{(2)}$ is higher 
than the wetting temperature $T_{\mathrm{w}}^{(1)}$ and for which the variation
in the thickness of the adsorbed layer is negligible at temperatures
close to $T_{\mathrm{w}}^{(1)}$. The advantange of doing 
is that one can more easily identify the origin of certain 
features of the surface phase behaviour (see below). 
The microscopic units of length of the problem
are given by $\lambda\!\equiv\!(A/\kappa)^{1/4}$. In the
calculation, we have fixed the range of the forces to be $p\!=\!2$
and the thickness $\ell_2^\pi\!=\!4\lambda$. The qualitative features
of the phase diagram described here are not in any way
specific to these values.\\

For $L_1\!=\!12\lambda$, the system undergoes an 
unbending transition represented by a continuous thick 
line ending in a critical point (see Fig.\ 3).
Along the line, the adsorbed layer on the stripe
can have either of two different thickness, both of them 
{\it thinner} than the surrounding adsorbed layer $\ell_2^\pi$. 
Note that the unbending line joins the liquid-vapour
coexistence line ($\Delta\mu\!=\!0$) at a certain 
non-zero angle. This is consistent with Eq.\ (\ref{clacla2})
and the fact that the difference in coverage 
$\Delta\widetilde{\Gamma}\!\sim\!\Delta\ell_1$ 
between the coexisting phases at that point is finite.
This transition has been predicted recently
\cite{Granada}, although the $T\!-\!\mu$ phase diagram had not been
described. Observe that the critical temperature of this transition
lies below $T_{\mathrm{w}}^{(1)}$, as predicted above. 
As the stripe width is increased, the length of the
unbending line decreases and for a given finite value 
of the width $L_1\!=\!L_1^{\dagger}\!\approx\!53\lambda$,
it vanishes (white circle in Fig.\ 3).
This phenomenology is {\it identical} 
to that described for a homogeneous but corrugated substrate \cite{RPS},
showing that the physics involved is essentially the same: the transition
takes place due to a balance between the free-energy associated with 
the direct intemolecular interaction with the substrate and the energetic
cost of increasing the area of the liquid-vapour interface. For this 
particular model, Eq.\ (\ref{eleven}), the transition vanishes for
\begin{equation}
\label{thirteena}
L_1^{\dagger}=\frac{2\kappa}{\,A\,(p\!+\!1)\,}
\left(\frac{\,(p\!+\!1)\;\ell_2^\pi\, }{p\!+\!3}\right)^{p+3}
\end{equation}
at a temperature
\begin{equation}
\label{thirteenb}
\Delta T^{\dagger} =
\frac{(p+3)^2}{\,p\,(p\!+\!1)\,}\,\frac{A}{\,\ell_2^\pi\,}.
\end{equation}

The loci of the critical points are represented in the same figure
by a grey thick line. For $L_1\!>\!L_1^{\dagger}$, 
this line extends further into the metastable
region of the phase diagram (dashed grey line) and, in the limit
$L_1\!\rightarrow\!\infty$, merges asymptotically with the wetting critical 
point at $T_{\mathrm{w}}^{(1)}$ (large black circle in Fig.\ 3). 
The presence of a (metastable)
critical point close to the coexistence line will affect the
adsorption properties in the {\it stable} part of the phase diagram.
For instance, the adsorption on the stripe will show a strong
increase close to that (metastable) critical point.

The loci of the points for which $\ell_1^\pi\!=\!\ell_2^\pi$ are
represented, in the same figure, by a straight black dotted line.
Whilst this line does not represent any surface phase transition
or singular behaviour, it does significantly influence the possible form
of the phase diagram. To see this, recall, as mentioned at 
the beginning of this section, that the adsorbed layer
is flat along that line for {\it any} value of the stripe width $L_1$.
Thus, neither the loci of the critical points nor
any unbending phase boundary can cross the line 
$\ell_1^\pi\!=\!\ell_2^\pi$.

\subsection{First-order Wetting Stripe}

We now suppose that the stripe corresponds to a material that
exhibits a first-order wetting transition in the limit of infinite
stripe size. For this case, we use an effective potential\cite{Review}

\begin{equation}
\label{nineteen}
W_1(\ell)=\left(\frac{A^2}{\,4\,B\,}\!+\!\Delta T\right)\frac{1}{\ell^{\,p}}-
\frac{A}{\,\ell^{\,p+1}}+\frac{B}{\,\ell^{\,p+2}}+\Delta\mu\,\delta\rho\;\ell,
\end{equation}
where $\Delta T\equiv T-T_{\mathrm{w}}^{(1)}$, $A$ and $B$ are (positive)
Hamaker constants and, as earlier, $p$ determines the range of the forces.
This potential models the discontinuous divergence of the film
thickness $\ell_1^{\pi}$ at $T_{\mathrm{w}}^{(1)}$, and an
associated pre-wetting line off coexistence for $\Delta T\!>\!0$.
This pre-wetting line ends in a pre-wetting critical point
$(\Delta T_{\mathrm{pw}},\Delta\mu_{\mathrm{pw}})$ where
\begin{equation}
\label{twentya}
\Delta T_{\mathrm{pw}}=\,\frac{1}{\,2\,p\,(p+3)\,}\;\frac{A^2}{B}
\end{equation}
\begin{equation}
\label{twentyb}
\Delta\mu_{\mathrm{pw}}\,\delta\rho=\,\frac{2\,B}{\,p+1\,}\left[
\frac{p+1}{\,2\,(p+3)\,}\,\frac{A}{B}
\right]^{\,p+3}.
\end{equation}
At this point, the wetting layer thickness is
\begin{equation}
\label{twentyc}
\ell_{\mathrm{pw}}=\,\frac{2\,(p+3)}{\,p+1\,}\,\frac{B}{A}
\end{equation}
which will be useful in the following analysis.

A necessary condition for criticality in the finite
stripe system is that the effective potential
has a point where the third derivative vanishes, (Eq.\ (\ref{tenb})). 
This provides an upper bound in temperature for any 
possible critical point in the system,
$\Delta T_C\!<\!3\,A^2/(4p(p\!+\!4)B)$. In contrast with the
previous case (section \ref{sec2}), a critical point can appear
in this system at temperatures above the wetting temperature 
$T_{\mathrm{w}}^{(1)}$ (but below a certain threshold). The conditions
for the critical point, Eqs.\ (\ref{nine})-(\ref{tenb}), 
read
\begin{equation}
\label{twentyone}
(p+1)\,A\,\ell_{1_{\scriptstyle C}}-2\,(p+3)\,B+
\frac{2\,\kappa}{L_1}\,\ell_{1_{\scriptstyle C}}^{\;p+4}=0,
\end{equation}
\begin{equation}
\label{twentytwo}
\Delta T_C = \frac{p+3}{p}\,\frac{A}{\ell_{1_{\scriptstyle C}}}
-\frac{(p+3)(p+4)}{p(p+1)}\,\frac{B}{\ell_{1_{\scriptstyle C}}^2}
-\frac{A^2}{\,4\,B\,},
\end{equation}
and
\begin{eqnarray}
\label{twentythree}
\Delta\mu_C\,\delta\rho=\frac{2\kappa}{L_1}\,\left[\;\ell_2^\pi
-\frac{p+3}{p+1}\,\ell_{1_{\scriptstyle C}}\right]+
\frac{2}{p+1}\,\frac{B}{\;\ell_{1_{\scriptstyle C}}^{\;p+3}\,}.
\end{eqnarray}

In the limit $L_1\!\rightarrow\!\infty$, these equations
recover (\ref{twentya})-(\ref{twentyc}), showing that the
critical point of the system must merge into the pre-wetting
critical point for large stripe widths, as expected
(see section \ref{sec1}). Notice that Eq.\ (\ref{twentyone})
implies that the thickness at the unbending critical point
$\ell_{1_{\scriptstyle C}}$ must be lower than the pre-wetting 
critical thickness $\ell_{\mathrm{pw}}$ for any finite value of $L_1$.
This allows us to sharpen our upper bound for the finite
stripe width critical temperature to $T_C\!<\!T_{\mathrm{pw}}$
for all widths $L_1$.
%These two properties are {\it independent} of
%the specific functional form of the effective potential 
%(in this case Eq.\ (\ref{nineteen})) and can be deducted
%geometrically from a generic effective potential describing
%a first-order wetting transition.

Figures 5-8 show the phase 
diagrams obtained for the effective
potential (\ref{nineteen}) and different values of the stripe
width $L_1$. Once again, we have chosen
$\lambda\!\equiv\!(A/\kappa)^{1/4}$
as unit of length and the range of the forces has
been fixed to $p\!=\!2$. Besides, due to the large number 
of parameters, we have fixed the Hamaker constant $B\!=\!\lambda A$.
As in the previous calculation, the
thickness of the adsorbed layer in the surrounding substrate,
$\ell_2^\pi$, has been considered constant (independent of
the temperature and the chemical potential).
However, in contrast to the case of critical wetting,
different values of $\ell_2^\pi$
lead to quite different surface phase behaviours.
As we shall show, the discriminating parameter here
is the ratio of $\ell_2^\pi$ to $\ell_{\mathrm{pw}}$.

The phase diagram of this system for $\ell_2^\pi\!=\!6\lambda$
and $L_1\!=\!2000\lambda$ is plotted in Fig.\ 5. As before, we find a
line of first-order undending phase transitions ending at a critical
point (black continuous line). Along this line, phases with
distinct adsorptions coexist (see insets). The line
$\ell_1^\pi\!=\!\ell_2^\pi$ represents the loci of points 
for which the adsorbed layer thickness is homogeneous even
though the substrate is heterogenous. The line serves to
separate interfacial configurations with different convexities \cite{BD3}. 
An extension of this line goes beyond the pre-wetting line
and joins the coexistence line. Along this extension (long dashed line), 
the liquid interface is also flat due to the presence of a metastable 
minimum or even a maximum of the effective potential (\ref{nineteen}) 
for $\ell\!=\!\ell_2^\pi$ \cite{Met}. Note that this scenario differs
from that occuring for a second-order wetting effective
potential. In that case,
the line $\ell_1^\pi\!=\!\ell_2^\pi$ never hits the coexistence line
for any value of the stripe width (see Fig.\ 3) and, therefore,
coexistence {\it always} takes place between adsorbed layers
thinner than the layer adsorbed on the surrounding substrate. 
In either case, we designate the transition as {\it unbending} 
since it originates from the same balance between the energy
associated to the interaction with the substrate and the 
energetic cost of increasing the area of the liquid-vapour 
interface \cite{Granada,RPS}. 

Observe also in Fig.\ 5 that the unbending line joins
the liquid-vapour
coexistence line ($\Delta\mu\!=\!0$) at a certain non-zero angle 
whilst the pre-wetting line approaches that line tangentially.
This is due to the fact that the change in the order 
parameter, $\ell_1$, across the unbending transition is {\it finite}
for the unbending transition while is infinite for the 
pre-wetting transition (because the system is wet for 
$T\!>\!T_{\mathrm{w}}^{(1)}$ at liquid-vapour coexistence)
[see Eq.\ (\ref{clacla2})].

Figure 6 shows the phase diagram of this system for 
$\ell_2^\pi\!=\!10\lambda$ and different values of the
stripe width $L_1$. As $L_1$ increases, the coexistence line 
approaches the pre-wetting line. A detailed description of 
the merging of both lines is given below. Notice the presence of the
line $\ell_1^\pi\!=\!\ell_2^\pi$ above the pre-wetting transition
and the fact that the unbending transition does not cross
that line for any value of $L_1$.

The phase diagram for $\ell_2^\pi\!=\!4\lambda$ and different 
values of the stripe width $L_1$ is presented in Fig.\ 7. As 
expected, the unbending line merges the pre-wetting line in
the limit $L_1\!\rightarrow\!\infty$. Note that the merging 
is {\it not} monotonic: as $L_1$ increases, the unbending line
crosses the pre-wetting line and merges from above, close to
the coexistence line, and from below, close to the pre-wetting
critical point (see inset). The locus of the unbending critical 
points is similarly non-monotonic for two reasons.
On one hand, for small values of $L_1$, the
unbending transition occurs at $(\Delta T,\Delta\mu$ 
where the potential (\ref{nineteen}) has
only one minimum. Consequently, the loci of the
unbending critical points resemble those shown
in Fig.\ 3 for critical wetting. 
On the other hand, for larger values of $L_1$, the
unbending transition takes place at $(\Delta T,\Delta\mu)$ 
where the potential (\ref{nineteen}) 
shows the characteristic double minimum of pre-wetting 
phase coexistence and the loci of unbending critical point
re-routes towards the pre-wetting critical point. The upshot
of this is that pre-wetting only has influence on the
surface phase diagram for sufficiently large stripe
widths $L_1$.

The contrast between these mechanisms is clearly 
magnified for thinner adjacent wetting layers.
Fig.\ 8 shows the phase diagram obtained for 
$\ell_2^\pi\!=\!\ell_{\mathrm{pw}}\!=\!10\lambda/3$.
Observe that the loci of the unbending critical points
now cross the liquid-vapour coexistence curve and, for a certain
range of width stripes 
($94\lambda\lesssim\!L_1\!\lesssim938\lambda$), the 
unbending phase transition is not present since it
is preempted by bulk condensation. The dissappearence
of the unbending transition as the stripe width 
increases (for $L_1\!\approx\!93.6\lambda$) resembles 
the behaviour for critical wetting (see Fig.\ 3).
In contrast, when $L_1\!\gtrsim937.5\lambda$,
the unbending transition {\it re-enters} the phase
diagram (at temperatures above $T_{\mathrm{w}}^{(1)}$)
and, in the limit $L_1\!\rightarrow\!\infty$, fuses
with the pre-wetting line. Notice once more that the
behaviour of the unbending line with the stripe width
is non-monotonic: after re-entering the phase diagram, 
its overall position first rises in temperature and,
for larger values of $L_1$, lowers towards the 
pre-wetting line (see inset). For the particular effective
potential (\ref{nineteen}), and within the context of the
minimal model, this re-entrant behaviour for unbending 
occurs when
\begin{eqnarray}
\label{twentyeigth}
\frac{\ell_2^\pi}{\ell_{\mathrm{pw}}}\;<\;\frac{(p+2)^2}{(p+1)(p+3)}.
\end{eqnarray}
Observe that the disappearence of the unbending transition
(and further re-entry at higher temperatures)
for small values of $\ell_2^\pi$ is inevitable if the
unbending transition is not to cross the iso-adsorption
line $\ell_1^\pi\!=\!\ell_2^\pi$, since this line 
surrounds the pre-wetting critical point as $\ell_2^\pi$
decreases (see Fig.\ 4). For this reason, this
feature must survive in analysis based on 
the full interfacial model, Eq.\ (\ref{two}).

\subsection{Discussion}
\label{Disc}

Before we turn to the results for a periodic array of
stripes, we complete this section with some critical remarks concerning 
the features of the phase diagrams predicted above.
To facilitate this, it will be instructive
to compare our results with those obtained within the context of the
full interfacial model (\ref{two}), by
Bauer and Dietrich (BD) for a particular stripe system made of 
a material which undergoes a first-order phase transition (when
infinite) \cite{BD3}. Specifically, let us concentrate on the
$T\!-\!\mu$ section of the phase diagram shown in Fig.\ 2 of their article. 
Whilst their diagram resembles our Fig.\ 7 in a 
qualitative way, there are three features which 
deserve detailed comment:
\begin{itemize}
\item In the limit $L_1\!\rightarrow\!\infty$, the adsorption
characteristics of the stripe must tend to those of the pure
substrate, {\it i.e.} the unbending line must merge with the pre-wetting 
line. Both approaches capture this requirement, although
they differ in the manner in which the mentioned lines coalesce.
The main difference lies in the behaviour 
of the unbending line close to liquid-vapour coexistence.
In the BD phase diagram, the unbending line fuses the
pre-wetting line monotonically from lower temperatures.
In contrast, as described above, the minimal model predicts
a non-monotonic merging: close to liquid-vapour coexistence,
the unbending line approaches the pre-wetting line from higher 
temperatures, as $L_1\!\rightarrow\!\infty$ (see, for example, 
the inset of Fig.\ 7). 

A detailed calculation with the full interfacial model (\ref{two}), 
included in appendix \ref{appA}, shows that either of both situations
is possible, and that the actual prevalence of one or the other
depends on a number of factors. However, for the experimentally
relevant case of dispersion forces, $p\!=\!2$, we believe the 
scenario described by the minimal model is the correct one:
the unbending line merges with the pre-wetting line from higher
temperature {\it close} to the liquid-vapour coexistence curve.
We suspect that numerical problems involved in the 
large-scale computation of BD are to blame here. For future
numerical studies, we note that the behaviour of the unbending line 
close to coexistence is sensitive to the range of the forces.
Consequently, the introduction of a cut-off in the {\it intermolecular 
potentials} may well reduce the effective range of the forces, 
producing a different result for the asymptotic behaviour
of the unbending line.

\item The intersection of the unbending coexistence line
with the bulk liquid-vapour coexistence line is tangential
in the BD phase diagram. This feature cannot be correct
because the difference in coexisting adsorptions remains
finite at $\Delta\mu\!=\!0$. 

\item The loci of the unbending critical point in the BD phase
diagram show a nontrivial behaviour as a function of the
stripe width $L_1$. That behaviour is analogous to that of
Fig.\ 7 and is explained above in terms of unbending in the 
presence or absence of energetic barrier in the effective potential
$W_1$. However, the BD phase diagram shows that, additionally, 
for even smaller values of the stripe width, the loci of unbending critical 
points curve towards the liquid-vapour coexistence line.
To explain that, we note that the minimal model prediction for
the critical temperature, Eq.\ (\ref{twentytwo}), does {\it not}
depend on the thickness of the adsorbed layer on the surrounding
substrate $\ell_2^\pi$, while the critical chemical potential,
Eq.\ (\ref{twentythree}), does (in fact, in very simple manner).
The curving of the loci of unbending critical points towards
the liquid-vapour coexistence line for very narrow stripes,
found by BD, is therefore a direct consequence of the {\it thinning} 
of the adsorbed layer on the surrounding substrate. 
This illustrates the merit of first keeping the value of 
$\ell_2^\pi$ constant before allowing for its own
dependence on temperature and chemical potential.

\end{itemize}

\section{Periodic Stripes}

\subsection{General Considerations}

Consider next a periodic array of stripes made of two different
materials (Fig.\ 1(b)) whose adsorption properties can be described, 
when pure, by the effective potentials $W_1(\ell)$ and $W_2(\ell)$,
respectively. The widths of the stripes are denoted $L_1$
and $L_2$, and the period $L\!=\!L_1\!+\!L_2$. This system
can be considered as the simplest prototype for studying
heterogenous wetting since for a single stripe the wetting
properties are completely determined by the embedding material. 
Thus, the periodic system can be used to study the dependence
of the wetting temperature on different factors such as composition,
degree of domain separation, etc.\ This has already been done to a
certain degree within density functional theory \cite{Frink}, 
but this approach seems more suitable to study packing phenomena 
close to the heterogeneities of the surface. As regards the
global wetting properties, we believe that a simpler approach
will suffice. This system has also been studied experimentally
\cite{Drelich}, mainly in the context 
of contact angles and Cassie's empirical law \cite{Cassie}.
This law states that the contact angle $\theta$ of a macroscopic 
drop placed on a planar heterogeneous substrate satisfies
\begin{equation}
\label{Cas}
\cos\theta = \sum_{i}\;\gamma_i\;\cos\theta_i,
\end{equation}
where $\theta_i$ is the contact angle of a droplet on a (pure) 
material of type $i$, and $\gamma_i$ is the fractional 
area of the substrate made of that material.
The implication of this law for the wetting temperature of a heterogeneous
system is straightforward:
\begin{equation}
\label{Cas2}
T_{\mathrm{w}} = \max\{\,T_{\mathrm{w}}^{(1)},T_{\mathrm{w}}^{(2)},\dots \},
\end{equation}
where $T_{\mathrm{w}}^{(i)}$ is the wetting temperature of the 
substrate $i$ (when pure). Here we check the validity of
this law with the help of the minimal model
put forward in section II. First, however, we recall briefly some general 
considerations concerning this system:

\begin{itemize}
\item For any given period $L$, the limiting cases $L_1\!\rightarrow\!0$ 
and $L_2\!\rightarrow\!0$ must produce the pure systems 1 and 2,
respectively. This fact follows trivially from the Hamiltonian
(\ref{five}). Observe, however, that it contradicts clearly 
Cassie's law since Eq.\ (\ref{Cas2}) states that 
infinitesimal amounts of a substance can change the wetting
properties of the embedding substrate.

\item If the thickness of the adsorbed layers on both infinite
substrates is the same ($\ell_1^\pi\!=\!\ell_2^\pi$), the functional
(\ref{two}) is fully minimised, term by term, by a flat solution
$\ell(x)\!=\!\ell_1^\pi\!=\!\ell_2^\pi$, as in
the single stripe system. For that reason, the loci of the points 
for which $\ell_1^\pi\!=\!\ell_2^\pi$ are not to be crossed by
any surface phase coexistence line. These are homogeneous
states on a heterogeneous substrate.
Note that since both $\ell_1^\pi$ and
$\ell_2^\pi$ vary with the temperature and chemical potential, 
the loci of the points for which $\ell_1^\pi\!=\!\ell_2^\pi$ 
may not be simple in the $T\!-\!\mu$ plane. Indeed,
the line may not exist at all.

\item If an unbending first-order transition takes place between 
different surface phases,
the coexistence line in the $T\!-\!\mu$ plane will verify 
the Clausius-Clapeyron-like equations (\ref{clacla}) and
(\ref{clacla2}).
\end{itemize}

\subsection{Minimal Model}

As described in section II, the free-energy of this system is given,
within the minimal model prescription, by the two-variable
function $H(\ell_1,\ell_2)$, Eq.\ (\ref{five}).
This free-energy depends on the {\it average} thicknesses $\ell_1$
and $\ell_2$, whose equilibrium value, for a given temperature
and chemical potential, is determined by minimisation of the 
free-energy. This yields the following coupled equations:
\begin{equation}
\label{thirty}
\begin{array}{ll}
2 \kappa\,(\ell_1-\ell_2)+L_1\,W_1'(\ell_1) & =\;0 \\
&\\
2 \kappa\,(\ell_2-\ell_1)+L_2\,W_2'(\ell_2) & =\;0.
\end{array} 
\end{equation}
Adding both equations we obtain the {\it sum rule}
\begin{equation}
\label{thirtyone}
L_1\,W_1'(\ell_1)\;+\;L_2\,W_2'(\ell_2)\;=\;0,
\end{equation}
which is the counterpart of another, obtained with the
full interfacial model (\ref{two})
\begin{equation}
\label{thirtytwo}
\int_{{\cal L}_1}\!\!\!dx\,\;W_{1}'(\ell)+
\int_{{\cal L}_2}\!\!\!dx\,\;W_{2}'(\ell)= 0.
\end{equation}
Eqns.\ (\ref{thirty}) can be solved by means of a
graphical construction. The solution 
corresponds to the intersection of the curves
\begin{equation}
\label{thirtythree}
\begin{array}{c}
\ell_2=\ell_1+\frac{L_1}{2\kappa}\,W_1'(\ell_1)\\
\\
\ell_1=\ell_2+\frac{L_2}{2\kappa}\,W_2'(\ell_2),
\end{array} 
\end{equation}
in the $\ell_1\!-\!\ell_2$ plane,
as seen in Fig.\ 9. As before, if there are multiple solutions,
the stable one can be discriminated upon comparison
of their free-energies, Eq.\ (\ref{five}).

Among these solutions, it is straightforward
to identify the spinodal points since they occur when the two 
curves of the graphical construction are tangential 
to each other at a given point. This leads 
to the following equation
\begin{equation}
\label{thirtyfive}
\left(1+\frac{L_1}{2\kappa}\,W_1'(\ell_1)\right)\;
\left(1+\frac{L_2}{2\kappa}\,W_2'(\ell_2)\right)=1,
\end{equation}
which can be obtained, alternatively, 
from the vanishing of the determinant of the Hessian of the free-energy 
$H(\ell_1,\ell_2)$. In addition, possible critical 
points can occur if (\ref{thirtyfive}) is satisfied
at a point of inflexion on any of the curves:
\begin{equation}
\label{thirtysix}
\begin{array}{c}
W_1'''(\ell_1)=0\\
\textrm{or}\hspace{8cm}\\
W_2'''(\ell_2)=0.
\end{array} 
\end{equation}
The existence of {\it two} alternative conditions for 
the presence of a critical point suggests that, in principle, 
there could be {\it two} different critical points in the 
$T\!-\!\mu$ phase diagram. This is quite different
to that obtained for the single
stripe system. In particular, as the limiting cases
$L_2\!=\!0$ and $L_1\!=\!0$ are each associated with 
one of the equations (\ref{thirtysix}), it follows
that there must be {\it two} unbending coexistence
lines which, in principle, could occur simultaneousy
in the $T\!-\!\mu$ phase diagram. All these features,
which emerge naturally from only elementary considerations
of the minimal model, would be much more difficult to
extract from the full interfacial description (\ref{two}).

\subsection{Cassie's Law}

Before applying the model to some specific examples,
we make connection
between Cassie's law and the minimal model. To do
that, we define the fractional area of material 1 
as $\gamma\!\equiv\!L_1/L$ for a given period $L$.
This allows us to write the free-energy of the system 
per unit length in the following way
\begin{equation}
\label{thirtyseven}
h(\ell_1,\ell_2)=\frac{\kappa}{L}\left(\ell_1\!-\!\ell_2\right)^{\,2}+
\gamma\,W_1(\ell_1) + (1\!-\!\gamma)\,W_2(\ell_2),
\end{equation}
which, in turn, enables us to change from a description in terms of
$L_1$ and $L_2$ into another in terms of $L$ and $\gamma$.
In this latter case, $L$ can be understood as a measure
of the overal size of the stripes, while $\gamma$ will be 
the fractional area of material 1. This is obviously true
as long as $\gamma$ is not too close to the limiting cases
$\gamma\!=\!0$ and $\gamma\!=\!1$. These limiting cases
give rise (trivially) to the pure substrates 2 and 1, respectively
(for any value of the period $L$).

Consider now, that the period $L$ is very large. This 
represents systems for which the heterogeneities extend
over macroscopic areas. In this case, we can neglect the first term of 
(\ref{thirtyseven}) and, consequently, the variables $\ell_1$ and $\ell_2$
decouple. Minimisation of the energy is achieved by $\ell_1\!=\!\ell_1^\pi$ 
and $\ell_2\!=\!\ell_2^\pi$ and the free-energy of the system, 
$\omega(T,\mu)\!\equiv\!\min\,h$, can be approximated
\begin{equation}
\label{thirtyeight}
\omega(T,\mu)\approx\gamma\,W_1(\ell_1^\pi) + (1\!-\!\gamma)\,W_2(\ell_2^\pi),
\end{equation}
which is but a different way of writing Cassie's law, Eq.\ (\ref{Cas}),
as generally accepted. The minimal model, therefore, recovers Cassie's
phenomenological law in the limit of macroscopic heterogeneities.

The other limiting case, corresponding to small values of $L$, represents
systems for which the heterogeneities are
of microscopic size. These systems are quasi-homogenenous
since the two materials are closely intermingled on the
surface. Recall that our model was not intended to work under
these assumptions. In spite of this, the limit $L\!\rightarrow\!0$
produces a sensible result. This occurs because the first term
of the free-energy (\ref{thirtyseven}) becomes exceptionally important
and has to be minimized fully yielding $\ell_1\!=\!\ell_2$; 
{\it i.e.} the thickness of the adsorbed layer is approximately
the same at every point of the substrate. The system behaves,
in this limit, as a homogeneous system whose adsorption
properties will be given by the following {\it averaged} 
effective potential
\begin{equation}
\label{thirtynine}
\overline{W}(\ell)\equiv\gamma\,W_1(\ell) + (1\!-\!\gamma)\,W_2(\ell).
\end{equation}
This description tallies with the spirit of the construction of the 
effective potentials as an integral of the solid-liquid and liquid-liquid
interactions over the entire system \cite{Israel}.

Therefore, the minimal model seems to describe the behaviour of the
heterogenous system in the two mentioned limits (macroscopic
and microscopic heterogeneities) in a physical way. We are fully
aware of the fact that such a broad description is achieved 
by smearing out some subtle details, and that the model will
not work under certain special circumstances. Nevertheless,
we stress that the model is put forward just as a versatile 
approximation to the full effective interfacial description,
Eq.\ (\ref{zero}).

\subsection{Examples}

As suggested above, we expect a very rich behaviour in the 
periodic system of parallel stripes. A full exploration of the
phenomenology of this system is a colossal task, well beyond
the scope of this paper. Therefore, we only report here the
calculation of the wetting temperature for two heterogeneous
systems as an example of the method.

We consider substrates undergoing first-order wetting 
transitions (when pure), whose effective potentials
will be modelled by
\begin{equation}
\label{fortya}
W_1(\ell)=\;\left(\frac{A_{1}^2}{\,4\,B_1\,}\!+\!(T-T_{\mathrm{w}}^{(1)})
\right)\frac{1}{\ell^{\,p}}-
\frac{A_1}{\,\ell^{\,p+1}}+\frac{B_1}{\,\ell^{\,p+2}},
\end{equation}
and
\begin{equation}
\label{fortyb}
W_2(\ell)=\left(\frac{A_{2}^2}{\,4\,B_{2}\,}\!+\!\alpha(T-T_{\mathrm{w}}^{(2)})
\right)\frac{1}{\ell^{\,p}}-
\frac{A_2}{\,\ell^{\,p+1}}+\frac{B_2}{\,\ell^{\,p+2}},
\end{equation}
Out of bulk liquid-vapour coexistence, the usual term 
$\Delta\!\mu\,\delta\!\rho\,\ell$ must be added, although
we do not consider this possibility here. The parameters
$A_{1}$, $B_1$, $A_{2}$, and $B_2$ are (positive) Hamaker constants
characterising the pertinent potentials. Notice that
a dimensionless constant $\alpha$ has to be included 
(multiplying the temperature in $W_2(\ell)$) to account
for differences in surface entropy between the two pure systems.
We use $\lambda\!\equiv\!(A_1/\kappa)^{1/4}$ as unit of length.

From this vast space of parameters we have chosen two examples,
both for long-range (dispersion) forces, $p\!=\!2$.
For each of them, we have calculated the wetting temperature of the
periodic system as a function of the fractional area of material 1,
$\gamma$, for different values of the period $L$. As mentioned 
above, the modification of the wetting temperature with the
composition of the substrate constitute a distinctive and 
important characteristic of these periodic systems
since, for a single stripe, the wetting temperature is
necessarily determined by the (infinitely wide) abutting substrate.
The results, for different choices of parameters,
are shown in Figs.\ 10 and 11. 

Let us concentrate first on Fig.\ 10.
As expected, the wetting temperature ranges between $T_{\mathrm{w}}^{(2)}$
(for $\gamma\!=\!0$) and $T_{\mathrm{w}}^{(1)}$ (for $\gamma\!=\!1$),
independently of the value of the period $L$. For small periods,
the wetting temperature interpolates almost
linearly between $T_{\mathrm{w}}^{(2)}$ and 
$T_{\mathrm{w}}^{(1)}$ (as a function of $\gamma$). This linearity 
is merely fortuitous. In general, $T_{\mathrm{w}}(\gamma)$  will be a
non-linear function joining the limiting cases $T_{\mathrm{w}}^{(2)}$ and 
$T_{\mathrm{w}}^{(1)}$. Within the minimal model,
the wetting temperature of the heterogeneous system can be calculated,
in this limit $L\!\rightarrow\!0$, as the wetting temperature of the
averaged potential $\overline{W}(\ell)$, Eq.\ (\ref{thirtynine}).
More specifically, for the model effective potentials (\ref{fortya})
and (\ref{fortyb}), we can write an analytical expresion for
the wetting temperature in the mentioned limit
\begin{eqnarray}
\label{fortyone}
T_{\mathrm{w}}(\gamma)\;=\;\frac{\gamma\,T_{\mathrm{w}}^{(1)}\!+
\!\alpha(1\!-\!\gamma)\,T_{\mathrm{w}}^{(2)}}{\gamma\!+\!\alpha(1\!-\!\gamma)}
\hspace*{.75cm}
\\ \nonumber
\\ \nonumber
\hspace*{.55cm}
-\frac{\gamma(1\!-\!\gamma)(A_1 B_2\!-\!A_2 B_1)^2}
{(\gamma\!+\!\alpha(1\!-\!\gamma))B_1 B_2(\gamma B_1\!+\!(1\!-\!\gamma)B_2)},
\end{eqnarray}
which is clearly non-linear.

As the period is increased, Cassie's law, Eq.\ (\ref{Cas2}), is recovered and
$T_{\mathrm{w}}(\gamma)\approx T_{\mathrm{w}}^{(1)}\!>\!T_{\mathrm{w}}^{(2)}$,
for almost every value of the fractional area $\gamma$. Deviations
occur only for small $\gamma$. These deviations become progressively
less important for large periods $L$, as expected.

The same quantity, calculated for a different system, is plotted 
in Fig.\ 11. In this case, the differences between substrates 1 and 2
are stonger: the thicknesses of the adsorbed layers on the pure
substrates become very dissimilar for temperatures 
$T\!\lesssim\!T_{\mathrm{w}}^{(2)}$ and so do their 
interfacial tensions (free-energies). In particular, the adsorbed layer
on substrate 1 is (in general) thicker than the corresponding layer
of substrate 2 (see Fig.\ 12) and its free energy is lower.
This introduces new phenomenology in the phase behaviour.
For small periods, $L\!=\!0.1\lambda$, the wetting temperature
of the periodic system resembles an {\it inversed}
Cassie's law since 
$T_{\mathrm{w}}(\gamma)\approx T_{\mathrm{w}}^{(2)}\!<\!T_{\mathrm{w}}^{(1)}$,
for most values of the fractional area $\gamma$. The fact that
the adsorption properties of the substrates differ considerably 
appears to have a decisive influence on the adsorption properties of the 
system in the quasi-homogeneous limit, $L\!\rightarrow\!0$. 
As we increase the value of $L$, we begin to recover the behaviour 
dictated by Cassie's law. On route to this, however, we uncover unexpected
richness. For the parameters used here, this first occurs
for $L\!\gtrsim\!0.2\lambda$, and is clearly visible for
$L\!=\!1.0\lambda$. 
The broken line in Fig.\ 11 corresponds to a (first-order)
unbending transition where adsorbed layers of different 
thicknesses coexist (similar to those occurring for a
single heterogeneous stripe). This coexistence finishes at an
unbending critical point (represented by a black circle)
as it did in the single stripe system. Of course, in the periodic
system, the unbending phase transition is associated with an 
adsorption jump on both domains. Fig.\ 12 shows the
thicknesses of the adsorbed layers $\ell_1$ and $\ell_2$ on substrates
1 and 2, respectively, as a function of the temperature for a
particular system ($L\!=\!10\lambda$, $\gamma\!=\!0.1$)
corresponding to
those wetting temperatures plotted in Fig.\ 11.
The thickness of the layers adsorbed on the pure substrates $\ell_1^\pi$ and 
$\ell_2^\pi$ are also shown for the sake of comparison. Observe
that for $(T-T_{\mathrm{w}}^{(1)})\lambda/A_1\!>\!-0.02$ the 
liquid interface has unbent and is essentially flat
($\ell_1\!\approx\!\ell_2$). Bearing this in mind, we refer
to the phase diagrams of Fig.\ 11 and notice 
(for instance, for $L\!=\!1.0\lambda$ or $10\lambda$) that 
the presence of the unbending transition modifies the
qualitative shape of curve representing the wetting temperature
$T_{\mathrm{w}}(\gamma)$. Specifically, at the point where the unbending
transition merges with the curve $T_{\mathrm{w}}(\gamma)$, the latter
has a kink. At the right of that point, $T_{\mathrm{w}}(\gamma)$
grows with a slope similar to the slope of the unbending
coexistence line, and tends to $T_{\mathrm{w}}^{(1)}$ very rapidly 
(as a function of $\gamma$). We can interpret this as an 
unbending-mediated wetting transition. In other words, at 
$T_{\mathrm{w}}(\gamma)$ the interface unbends to the wet configuration 
$\ell_1\!=\!\ell_2\!=\!\infty$. On the other hand,
to the left of that merging point, the interface finds a stable
(unbent) flatter configuration ($\ell_1\!\approx\!\ell_2$) resembling 
those of the quasi-homogenenous limit ($L\!\approx\!0$). Observe
that $T_{\mathrm{w}}(\gamma)$ hardly changes by increasing $L$ 
in that region of the phase diagram from the curve obtained
for $L\!=\!0.1\lambda$. This region, and the unbending line,
are squeezed towards small values of the fractional area $\gamma$
for larger values of the period $L$. Notice that the unbending 
line can neither 
dissappear (because that limit contains the unbending transition
for the single stripe line) nor merge with the line $\gamma\!=\!0$
(representing pure substrate 2, which does not undergo any 
unbending transition). However, it becomes 
progressively and proportionaly less 
important as $L$ grows and remains confined to the region
of small fractional area, $\gamma\!\approx\!0$. In the
limit of macroscopic heterogeneities, $L\!\rightarrow\!\infty$, 
Cassie's law is recovered as discussed earlier.

\section{Conclusions}

In this paper, we have forwarded a simple model of surface
phase behaviour for fluid adsorption on heterogenenous micropatterned
substrates. This has been applied to two proto-typical systems: a)
a single stripe of a material exhibiting either a first- or 
second-order wetting transition embedded in an infinite partially
wet substrate and b) a periodic array of stripes each of which
exhibit first-order wetting at distinct temperatures. 
The advantage of our "minimal" description is that it allows us 
to explore the vast phase space of parameters that these systems
present. This would be an extremely arduous task within even an
effective interfacial Hamiltonian description, let alone a
microscopic density functional approach.
Despite the simplicity of our model, we rederive
results obtained previously using more microscopic methods for
specific choices of interfacial binding potentials. This gives us
confidence concerning the validity of our method. Applying the
minimal model to other choices of substrate heterogeneity, we
predict new examples of surface phase behaviour. In particular, 
for the single stripe system with first-order wetting, we establish 
the conditions under which the unbinding coexistence line approaches
the pre-wetting line (of the pure substrate) from above or below, in
the limit of large stripe widths. For the periodic system, we have
concentrated on establishing the value of the wetting transition temperature
as a function of the fractional area of the two materials as well as
the total period of the heterogeneity. Agreement with Cassie's
empirical law is obtained in the limit of macroscopically large
domain sizes. Thus, the wetting temperature of the heterogeneous
substrate is essentially determined by the maximum wetting temperature of
each of the pure components. For mesoscopic systems, on the other
hand, notable deviations from Cassie's law are possible. This includes 
"inversions" of Cassie's law where we observe that the wetting temperature 
of the heterogeneous substrate is close to the lower of the two 
wetting temperatures of the pure components. The value of the wetting
temperature may also be sensitive to the presence of unbending phase
transitions induced by the heterogeneous substrate. It is clear to
us that even within the context of the present minimal description,
the possible surface phase behaviour on the periodic micropatterned
substrate is extraordinarily rich. For example, out of bulk two
phase coexistence, additional phenomenology may arise due to the
competition between lengthscales associated with the pre-wetting
line on each pure component. A full exploration of the possible
surface phase behaviour requires much further work along these lines.

\vspace*{.5cm}

We are very grateful to Prof.\ M.E.\ Cates for support during the 
completion of this project.
C.R.\ acknowledges economical support from the E.C.\
under contract ERBFMBICT983229.\\

$^\star$ Present address: Dept.\ of Physics and Astronomy,
University of Edinburgh, Edinburgh EH9 3JZ, United Kingdom.

\begin{appendix}
\section{}
\label{appA}

Consider an infinite stripe of width $L_1$ made of material 1 
that is embedded in a substrate made of material 2 (Fig.\ 1(a)).
If these materials are such that their adsorption properties
can be described (when pure) by effective potentials $W_1$ and 
$W_2$, the adsorption properties of the combined system can
be obtained by the functional minimisation of the free-energy
(\ref{two}). If substrate 1 undegoes a first-order wetting 
transition at $T\!=\!T_{\mathrm{w}}^{(1)}$ when pure 
(and, therefore, its phase diagram
presents a pre-wetting line), the combined system must undergo 
an unbending transition whose coexistence line must merge with the 
pre-wetting line of the pure substrate in the limit 
$L_1\!\rightarrow\!\infty$ (as explained in the main text of the paper).

In this appendix, we want to determine under which circumstances
the interfacial model (\ref{two}) predicts that, in the mentioned
limit $L_1\!\rightarrow\!\infty$, 
the unbending line approaches the pre-wetting line from above
or from below ({\it i.e.} from temperatures higher or lower than the 
wetting temperature $T_{\mathrm{w}}^{(1)}$, respectively).
We denote these two possibilities as scenarios A (above) and
B (below).
As we are mainly interested in the merging of the
two lines close to the bulk liquid-vapour coexistence line
(see \ref{Disc}), we only need to concentrate on $\Delta\mu\!=\!0$.

To start, we notice that our goal can be achieved, without
lack of generality, by studying the system at the wetting 
point $T\!=\!T_{\mathrm{w}}^{(1)}$. Scenario A will be
correct if the stable configuration, at that point, 
corresponds to a bound state in the limit 
$L_1\!\rightarrow\!\infty$. On the other hand,
if the stable configuration is unbound in that limit, the
scenario B will be followed. In other words, for 
$T\!=\!T_{\mathrm{w}}^{(1)}$ and $\Delta\mu\!=\!0$,
there must be two different configurations which 
minimise the Hamiltonian (\ref{two}) in the limit 
$L_1\!\rightarrow\!\infty$ (see below). This occurs 
due to the presence of a local maxima (the activation barrier)
in the potential $W_1(\ell)$ at the wetting point. 
We denote these configurations by $\ell^{-}(x)$ (bound) and 
$\ell^{+}(x)$ (near-unbound). If $\ell^{-}(x)$ is the
stable configuration (it has lower energy), scenario
A will be correct. Otherwise, it will be scenario B.

Minimization of (\ref{two}) yields the following
Euler-Lagrange equation:
\begin{equation}
\label{app1}
\Sigma\;\ddot{\ell}\,(x) = \left\{
\begin{array}{ll}
W_1'(\ell) & \;\textrm{if }\, |x|\le L_1/2\\
& \\
W_2'(\ell) & \;\textrm{if }\, |x|>L_1/2,
\end{array} \right.
\end{equation}
where we have located the origin $x\!=\!0$ in the
center of the stripe. This
equation can be integrated once to yield:
\begin{equation}
\label{app2}
\frac{\Sigma}{2}\;\dot{\ell}\,(x)^{2} = \left\{
\begin{array}{ll}
\Delta W_1(\ell) -{\cal P}& \;\textrm{if }\, |x|\le L_1/2\\
& \\
\Delta W_2(\ell) & \;\textrm{if }\, |x|>L_1/2,
\end{array} \right.
\end{equation}
where $\Delta W_1(\ell)\!\equiv\!W_1(\ell)\!-\!W_1(\ell_1^\pi)$,
$\Delta W_2(\ell)\!\equiv\!W_2(\ell)\!-\!W_2(\ell_2^\pi)$, and
${\cal P}$ is a (non-negative) constant of integration to be determined. Note 
that the boundary condition at $x\!=\!\infty$ has been imposed. 
Further conditions read
\begin{equation}
\label{app3}
\begin{array}{lcc}
\Delta W_1(\ell_0) &\;=\;& {\cal P}\\
\\
\Delta W_1(\ell_L) &\;=\;& \Delta W_2(\ell_L) + {\cal P},
\end{array}
\end{equation}
where $\ell_0\!\equiv\!\ell(0)$ and $\ell_L\!\equiv\!\ell(L_1/2)$.
These two equations can be solved by means of a geometrical
construction for an arbitrary value of $\cal P$ (see Fig.\ 13).
The relevant value for our case (the limit $L_1\!\rightarrow\!\infty$) 
is ${\cal P}\!\rightarrow\!0$ (see below). Notice in the figure
that the different pairs of values $\ell_0^{-}$, $\ell_L^{-}$ 
and $\ell_0^{+}$, $\ell_L^{+}$ will give rise to the two 
solutions $\ell^{-}(x)$ and $\ell^{+}(x)$.

Here we can make our first prediction with the help of the 
geometrical construction: if $\ell_1^\pi\!\ge\!\ell_2^\pi$,
the correct scenario must be A for any physical shape
of the effective potentials $W_1$ and $W_2$. This happens because,
under these circumstances, {\it every} term of the Hamiltonian 
turns out to be lower for the bound configuration $\ell^{-}(x)$ 
than for the near-unbound one (if the latter exists at all). This 
(purely geometrical) feature is captured correctly by the minimal 
model (\ref{six}). However, if $\ell_1^\pi\!<\!\ell_2^\pi$
at the wetting point, the two configurations exist
(in this case $\ell^{+}(x)$ always exists for sufficientely 
large values of $L_1$) and their free-energies need to be calculated.

To proceed, we use equation (\ref{app2}) to write the quadrature
\begin{equation}
\label{app4}
L_1\;=\;\sqrt{2\Sigma}\;\left|\,\int_{\ell_0}^{\,\ell_L}\!\!
\frac{dx}{\sqrt{\Delta W_1(\ell)\!-\!{\cal P}}}\;\right|,
\end{equation}
which determines ${\cal P}$ implicitly as a function of the stripe
width $L_1$, $T$ and $\mu$. 
A simple calculation shows that ${\cal P}$ is 
related to the free-energy of the system in the following way:
\begin{equation}
\label{app7}
{\cal P} = \frac{\partial\,\omega}{\partial L_1} + 
W_2(\ell_2^\pi) - W_1(\ell_1^\pi).
\end{equation}
Thus, the comparison of the energies of
configurations $\ell^{-}(x)$ and $\ell^{+}(x)$
can be obtained from the asymptotic behaviour of
${\cal P}$ for large values of $L_1$.
In turn, this follows from studying the
asymptotic behaviour of (\ref{app4}) when
${\cal P}\!\rightarrow\!0$, since this is the
only value for which the integral diverges and recovers, 
in that way, the limit $L_1\!\rightarrow\!\infty$. For
appropriate effective potentials $W_1$ and $W_2$
(with the only requirement that $W_1$ has an activation
barrier), we obtain
\begin{equation}
\label{app8}
\begin{array}{cl}
{\cal P}^{-}\;\sim & L_1^{-4} \\
\\
{\cal P}^{+}\;\sim & L_1^{-\frac{\scriptstyle 2p}{\,\scriptstyle p+3\,}},
\end{array}
\end{equation}
where $(-)$ and $(+)$ denote the bound and near-unbound configuration,
respectively, and $p$ is the range of the forces, {\it i.e.} 
$W_1(\ell)\!\sim\!\ell^{-p}$ when $\ell\!\rightarrow\!\infty$.
Therefore, the energy of each configuration can be written
asymptotically as
\begin{equation}
\label{app9}
\begin{array}{cl}
\omega^{-}\;\sim & \omega_0^{-} +\,\omega_1^{-}\;L_1^{-3}\,+\dots\\
\\
\omega^{+}\;\sim & \omega_0^{+} +
\,\omega_1^{+}\;L_1^{\,\frac{\scriptstyle 3-p}{\,\scriptstyle p+3\,}}\,+\dots,
\end{array}
\end{equation}
where $\omega_0^{-}$, $\omega_0^{+}$, $\omega_1^{-}$ and $\omega_1^{+}$ are
constants that depend on the details of the potential. Note that the
irrelevant common term $\left[W_2(\ell_2^\pi)\!-\!W_1(\ell_1^\pi)\right]\,L_1$
has been subtracted. 

Consequently, the range of the forces $p$ appears to be 
determinant in establishing which of the two scenarios holds. For
$p<3$, the energy of the bound configuration tends to a constant
for large stripe widths, $\omega^{-}\!\rightarrow\!\omega_0^{-}$,
while the energy of the near-unbound configuration diverges as a
function of the stripe width $L_1$. For that reason,
the bound configuration $\ell^-(x)$ is the stable one in this case,
and scenario A is correct. This includes the experimentally 
relevant case of dispersion forces, $p\!=\!2$.
In contrast, for $p\ge3$, both $\omega^{-}$ and $\omega^{+}$
tend to constant values in the limit $L_1\!\rightarrow\!\infty$.
The prevalence of scenario A or B will be determined, in this latter
case, by the comparison of the free-energies:
\begin{equation}
\label{app10}
\frac{\omega_0^{-}}{2\sqrt{2\Sigma}}=
\int_{\ell_1^\pi}^{\,\ell_L}\!\!\!\!\!\!d\ell\;
\sqrt{\Delta W_1(\ell)}\,+
\int_{\ell_L}^{\,\ell_2^\pi}\!\!\!\!\!\!d\ell\;
\sqrt{\Delta W_2(\ell)},
\end{equation}
and
\begin{equation}
\label{app11}
\frac{\omega_0^{+}}{2\sqrt{2\Sigma}}=
\int_{\ell_L}^{\,\infty}\!\!\!\!\!\!d\ell\;
\sqrt{\Delta W_1(\ell)}\,+
\int_{\ell_2^\pi}^{\,\ell_L}\!\!\!\!\!\!d\ell\;
\sqrt{\Delta W_2(\ell)}.
\end{equation}

The necessary conditions for scenarios A or B to occur
are gathered in the following table:

\renewcommand{\arraystretch}{1.2}
\begin{center}
\begin{tabular}{|c|c|c|c|} \hline
\multicolumn{3}{|c|}{\hspace{1.5cm}{\sf Conditions}\hspace*{1.5cm}}
&\multicolumn{1}{|c|}{\hspace{1cm}{\sf Scenario}\hspace*{1cm}} \\ \hline \hline
\multicolumn{3}{|c|}{} &\multicolumn{1}{|c|}{} \\
\multicolumn{3}{|c|}{$\ell_1^\pi\!\ge\!\ell_2^\pi$} &\multicolumn{1}{|c|}{A} \\
\multicolumn{3}{|c|}{} &\multicolumn{1}{|c|}{} \\ \hline
 & \multicolumn{2}{c|}{$p<3$ } & A \\
\cline{2-4}
$\;\;\;\ell_1^\pi\!<\!\ell_2^\pi\;\;\;$  &  & $\omega_0^{-}\!<\!\omega_0^{+}$ &A \\
\cline{3-4}
 &\raisebox{2.25mm}[0pt]{$\;p\ge3\;$} & $\omega_0^{-}\!>\!\omega_0^{+}$ & B \\ \hline
\end{tabular}
\end{center}      

where $\ell_1^\pi$ and $\ell_2^\pi$ are the wetting layer thicknesses
on the pure substrates measured at $T\!=\!T_{\mathrm{w}}^{(1)}$.
\end{appendix}

\begin{figure}[t]
%\vspace*{3cm}  
\centerline{\epsfig{file=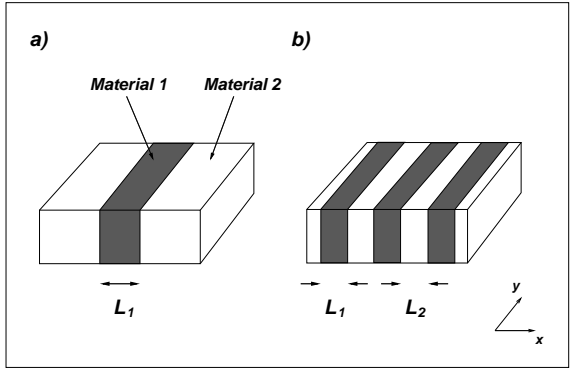,width=7.5cm}}
\vspace*{.75cm}  
\caption{Schematic illustration of micropatterned
substrates considered in this paper. The figure shows
a) a single stripe of width $L_1$ 
embedded in substrate 2 and b) a periodic array of stripes of widths
$L_1$ and $L_2$, made of materials 1 and 2, respectively.
Both systems are translationally invariant along the $y$ direction.}
\end{figure}

\begin{figure}[t]
%\vspace*{3cm}  
\centerline{\epsfig{file=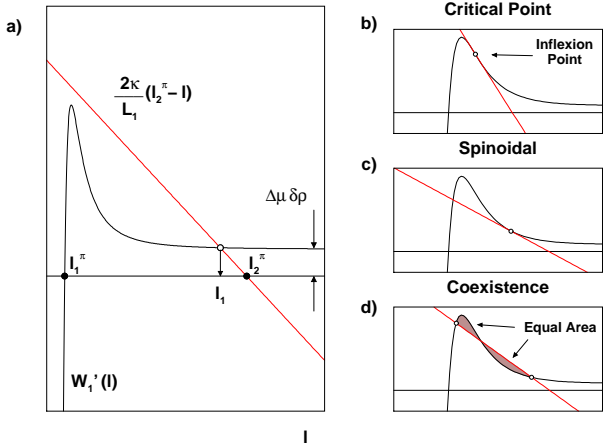,width=8.5cm}}
\vspace*{.1cm}  
\caption{Sketch of the graphical construction used to
solve Eq.\ (\protect\ref{nine}). The {\it average}
thickness of the adsorbed layer on the stripe, $\ell_1$, is obtained
by the intersection of the curve $W_1'(\ell)$ with the
straight line $2\kappa(\ell_2^{\pi}\!-\!\ell)/L_1$ (see (a)).
The geometrical conditions for a critical point (b), a
spinodal (c), and the coexistence between two different
values of the adsorption $\ell_1$ (d) are also shown.}
\end{figure}

\begin{figure}[t]
%\vspace*{3cm}  
\centerline{\epsfig{file=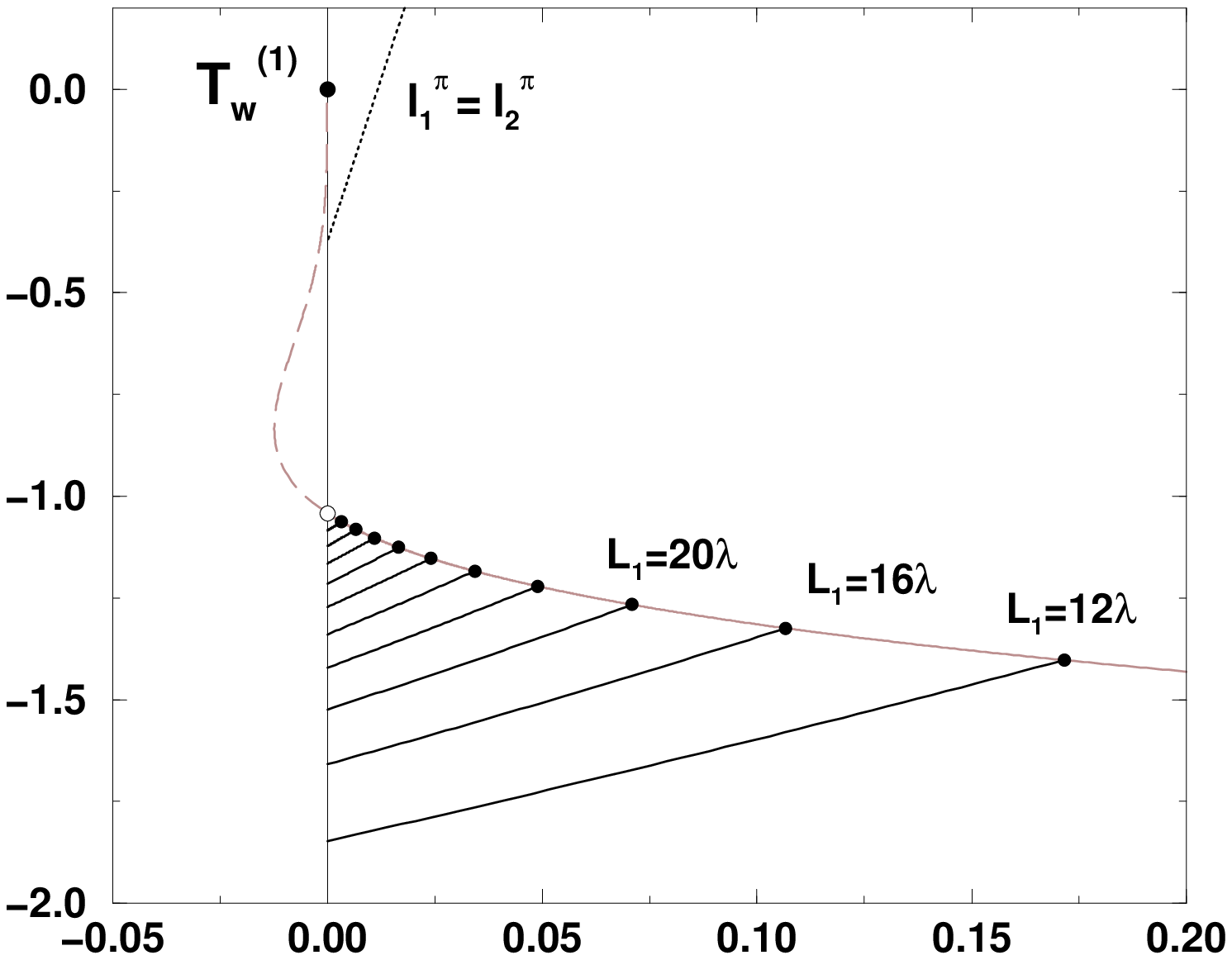,width=8.5cm}}
\vspace*{.5cm}  
\caption{$T\!-\!\mu$ phase diagram for the adsorption on a stripe
of different widths $L_1$. The stripe is made of a material that,
when pure, undergoes a {\it second-order} wetting transition at 
$T\!=\!T_{\mathrm{w}}^{(1)}$ (large full circle), 
whose adsorption properties are described by the effective potential
(\protect\ref{eleven}). The stripe is embedded in a partially wett
substrate with finite film thickness $\ell_2^{\pi}\!=\!4\lambda$.
This system shows an {\it unbending}
transition (black continuous lines) for a range of stripe widths.
In the figure, $L_1/\lambda\!=\!12,16,20,24,28,32,36,40,44$ and $48$.
For $L_1\!\approx\!53\lambda$, the unbending transition is pre-empted
by the bulk liquid-vapour phase transition (large empty circle). 
The continuous grey line represents the loci of the critical points 
(as given by Eqs.\ (\ref{twelvea}-\ref{twelvec}). This line
extends into the metastable region of the phase diagram 
(dashed grey line) and tends asymptotically to the 
wetting temperature $T_{\mathrm{w}}^{(1)}$ 
when $L_1\!\rightarrow\!\infty$. Note that the unbending transition
does {\it not} cross the iso-adsorption line $\ell_1^{\pi}\!=\!\ell_2^{\pi}$,
(dotted line).}

\begin{figure}[t]
%\vspace*{3cm}  
\centerline{\epsfig{file=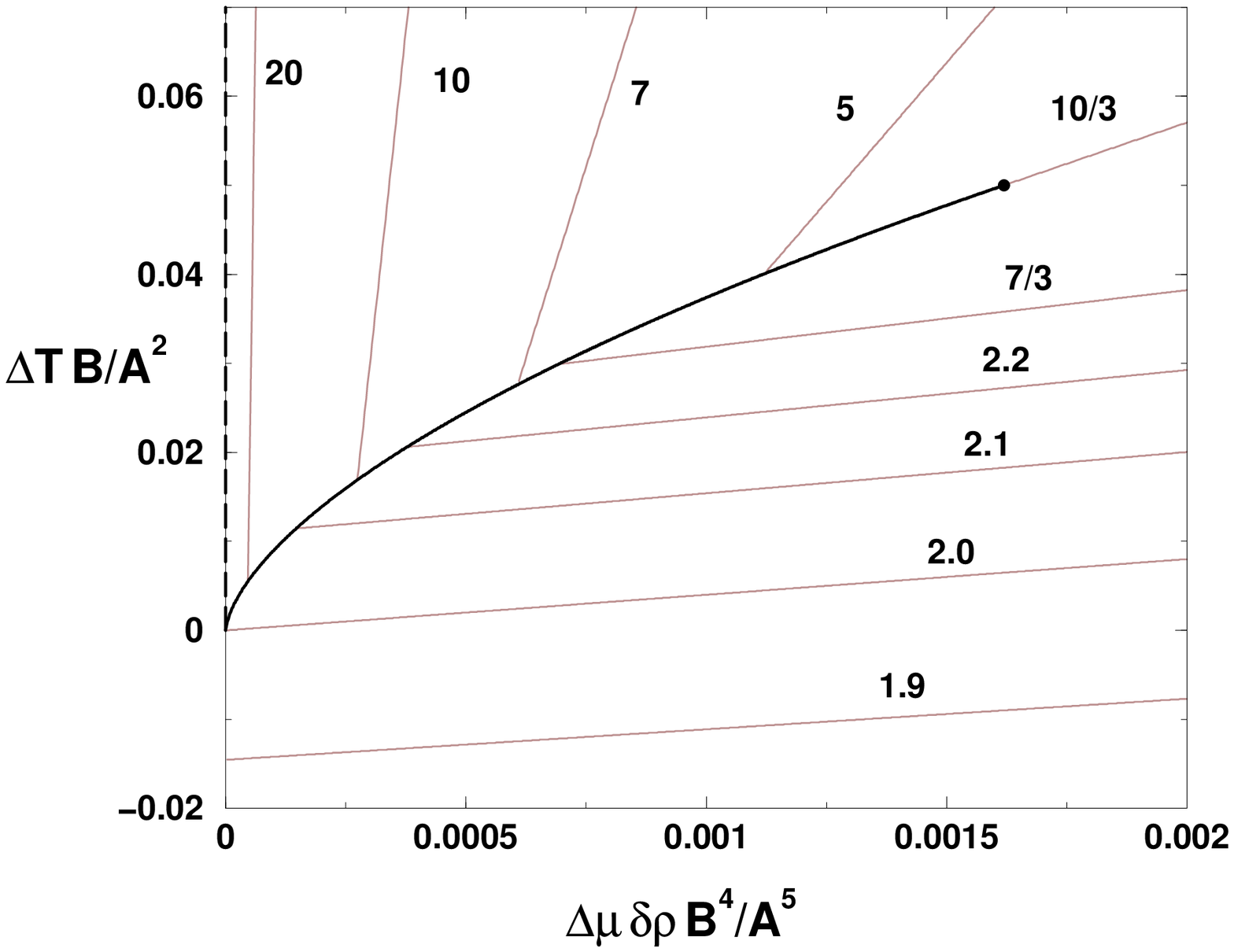,width=8.cm}}
\vspace*{.2cm}  
\caption{pre-wetting line (black) as modelled by the effective potential
(\protect\ref{nineteen}) with $p\!=\!2$. 
Iso-adsorption lines (grey) are also shown for different
values of the adsorption $\ell$ ranging between 1.9 and 20
(in units of $B/A$).}
\end{figure}

\end{figure}

\begin{figure}[t]
%\vspace*{3cm}  
\centerline{\epsfig{file=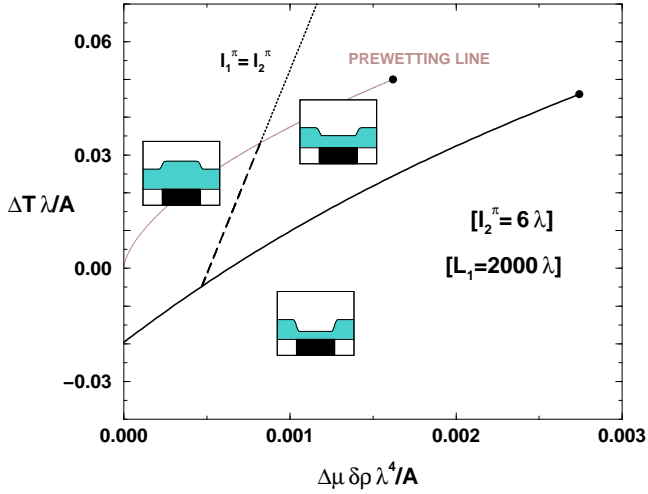,width=8.5cm}}
\vspace*{.5cm}  
\caption{$T\!-\!\mu$ phase diagram for the adsorption on a stripe
of width $L_1$ made of a material exhibiting a {\it first-order}
wetting transition (when pure). The stripe is embedded in a partially 
wet substrate whose wetting layer thickness $\ell_2^{\pi}$ is considered
independent of temperature and chemical potential. The pre-wetting
line of the (pure) material 1 is shown for comparison
(See text for details).}
\end{figure}

\begin{figure}[t]
%\vspace*{3cm}  
\centerline{\epsfig{file=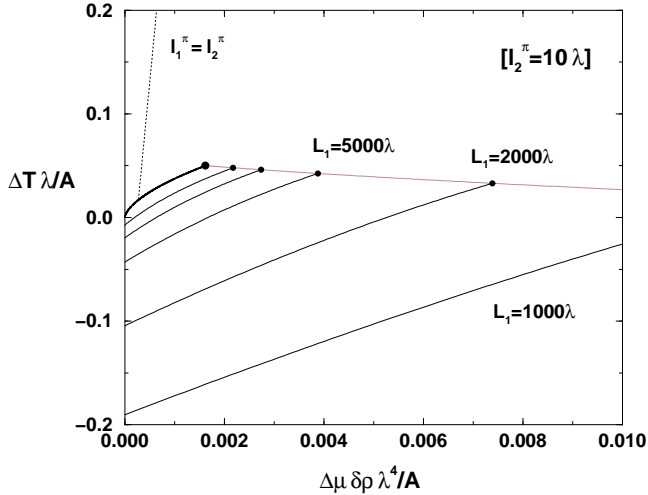,width=8.5cm}}
\vspace*{.5cm}  
\caption{$T\!-\!\mu$ phase diagram showing unbending phase transitions
occuring on a substrate with a single stripe showing a {\it first-order} 
wetting transition 
(when pure), embedded in a substrate whose wetting layer 
thickness $\ell_2^{\pi}\!=\!10\lambda$. The unbending coexistence
lines are represented by thin black lines which finish at
critical points, the loci of which is shown as grey. 
Different stripe widths
are considered: $L_1\!=\!1,2,5,10$ and $20$ ($\times10^3\lambda$).
The pre-wetting line of the (pure) material 1, which
corresponds to the unbending line in the limit 
$L_1\!\rightarrow\!\infty$, is shown for 
comparison (thick line). The broken line represents the line
for which $\ell_1^\pi\!=\!\ell_2^\pi$. Observe that this line is 
{\it not} crossed by any unbending line (See text for details).}
\end{figure}

\begin{figure}[t]
%\vspace*{3cm}  
\centerline{\epsfig{file=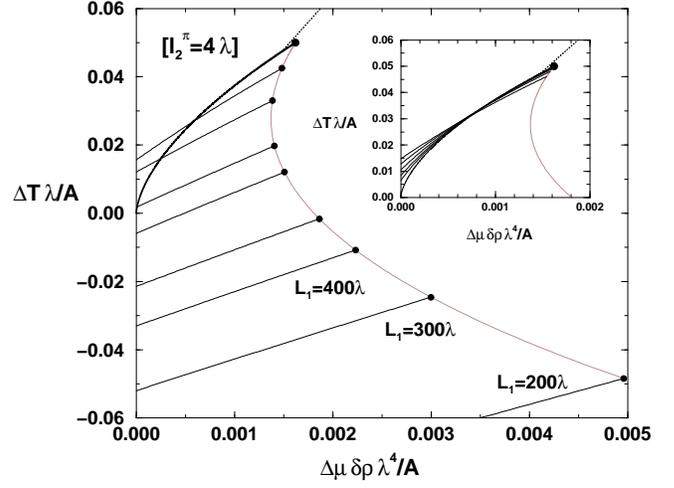,width=8.5cm}}
\vspace*{.5cm}  
\caption{As Fig.\ 6, with $\ell_2^{\pi}\!=\!4\lambda$ and
$L_1\!=\!2,3,4,5,7.5,10,20$ and $50$ ($\times10^2\lambda$).
Inset: Merging of the unbending line with the pre-wetting
line for large stripe widths, 
$L_1\!=\!1,2,4,8$ and $16$ ($\times10^4\lambda$).}
\end{figure}

\begin{figure}[t]
%\vspace*{3cm}  
\centerline{\epsfig{file=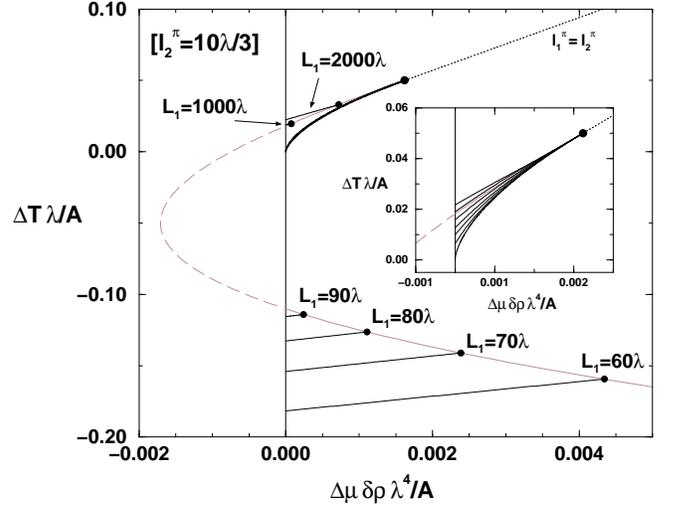,width=8.5cm}}
\vspace*{.5cm}  
\caption{As Fig.\ 6, with $\ell_2^{\pi}\!=\!10\lambda/3$ and
different values of $L_1$. The grey broken line is the
loci of the unbending critical points that have been pre-empted
by the bulk liquid-vapour phase transition.
Inset: Merging of the unbending line with the pre-wetting
line for large stripe widths, 
$L_1\!=\!4,8,16,32$ and $64$ ($\times10^3\lambda$).}
\end{figure}

\begin{figure}[t]
%\vspace*{3cm}  
\centerline{\epsfig{file=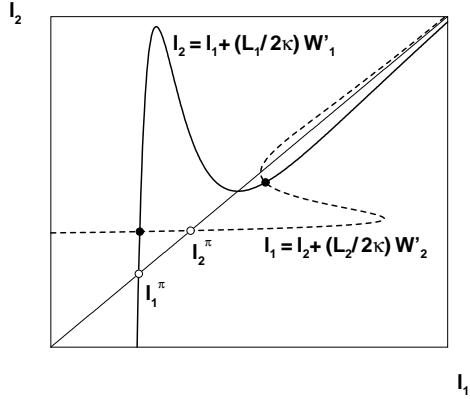,width=7.cm}}
\vspace*{.25cm}  
\caption{Sketch of the graphical construction used to
solve equations (\protect\ref{thirty}). In the figure,
both potentials $W_1$ and $W_2$ represent first order 
wetting transitions.
The solutions correspond to the intersection of the curves 
$\ell_2\!=\!\ell_1\!+\!\frac{L_1}{2\kappa}W_1'(\ell_1)$ and
$\ell_1\!=\!\ell_2\!+\!\frac{L_2}{2\kappa}W_2'(\ell_2)$
(black circles). If there are more than one, as in the
figure, the stable one will have the lower free energy, 
Eq.\ (\ref{five}). The thickness of the adsorbed layers on
the pure substrates $\ell_1^\pi$ and $\ell_2^\pi$
correspond to a crossing of the curves with the
line $\ell_1\!=\!\ell_2$ (white circles).}
\end{figure}

\begin{figure}[t]
%\vspace*{3cm}  
\centerline{\epsfig{file=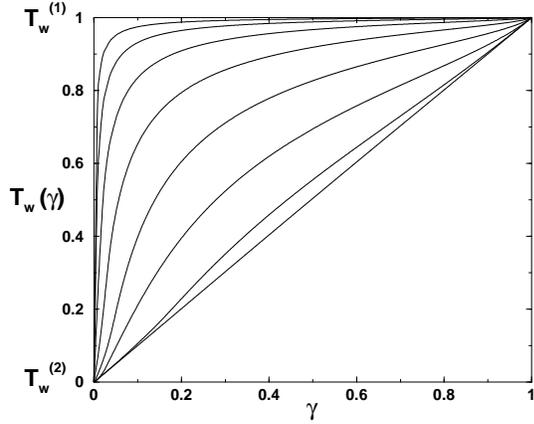,width=7.cm}}
\vspace*{.25cm}  
\caption{Wetting temperature (measured with respect to $T_{\mathrm{w}}^{(2)}$)
of a periodic heterogeneous substrate calculated
with the minimal model, Eq.\ (\protect\ref{five}) and the
effective potentials (\protect\ref{fortya}) and (\protect\ref{fortyb}).
The wetting temperature (in units of $A_1/\lambda$) is
represented as a function of the fractional area 
of the first material, $\gamma$, for different values of the
period $L/\lambda\!=\!1,10,10^2,10^3,10^4,10^5,10^6$ and $10^7$
(from below).
We use the following parameters: $T_{\mathrm{w}}^{(1)}\!-\!T_{\mathrm{w}}^{(2)}
\!=A_1/\lambda$, $B_1\!=\!2B_2\!=\!A_1\lambda$, $A_1\!=\!A_2$ and 
$\alpha\!=\!2$.}
\end{figure}

\begin{figure}[t]
%\vspace*{3cm}  
\centerline{\epsfig{file=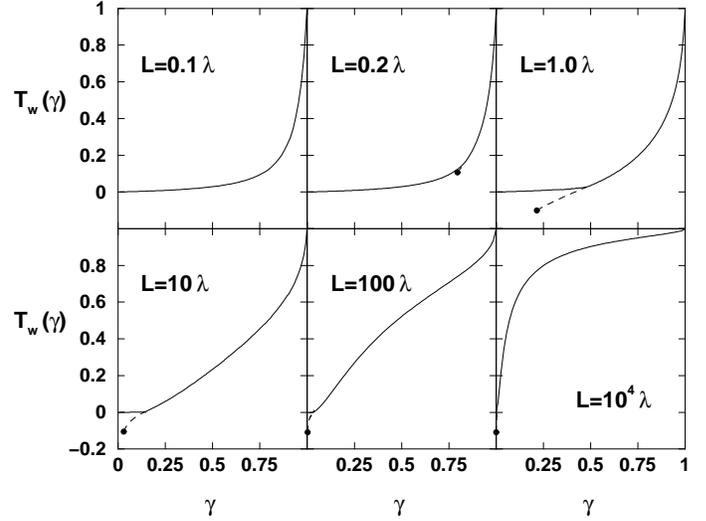,width=9.cm}}
\vspace*{.25cm}  
\caption{As Fig.\ 10, for 
$T_{\mathrm{w}}^{(1)}\!-\!T_{\mathrm{w}}^{(2)}\!=A_1/\lambda$,
$B_2\!=\!4\,B_1\!=\!2\,A_1\lambda$, $A_2\!=\!0.05\,A_1$
and $\alpha\!=\!20$. The broken line represents an
unbending coexistence line appearing for $L\!\approx\!0.2\lambda$
(see text for details).}
\end{figure}

\begin{figure}[t]
%\vspace*{3cm}  
\centerline{\epsfig{file=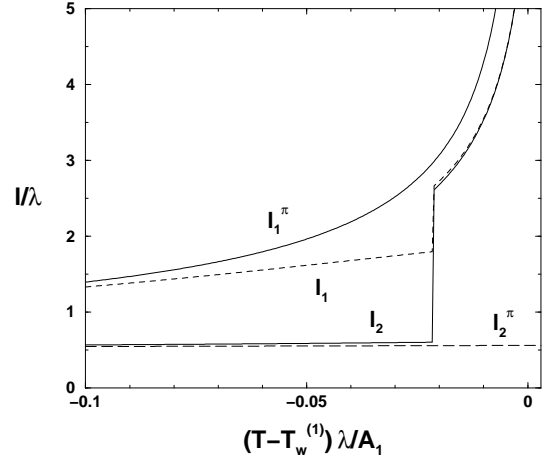,width=7.cm}}
\vspace*{.25cm}  
\caption{Adsorption on the periodic system whose wetting temperature 
is plotted in Fig.\ 11 for $L\!=\!10\lambda$ and $\gamma\!=\!0.1$.
The curves $\ell_1$ and $\ell_2$ represent the adsorption on
the substrates 1 and 2 (respectively). A (first-order) 
unbending transition takes place for 
$(T-T_{\mathrm{w}}^{(1)})\lambda/A_1\!\approx\!-0.02$.
For the sake of comparison,
the adsorption on the pure substrates 1 and 2 is also shown
($\ell_1^\pi$ and $\ell_2^\pi$).}
\end{figure}

\begin{figure}[t]
%\vspace*{3cm}  
\centerline{\epsfig{file=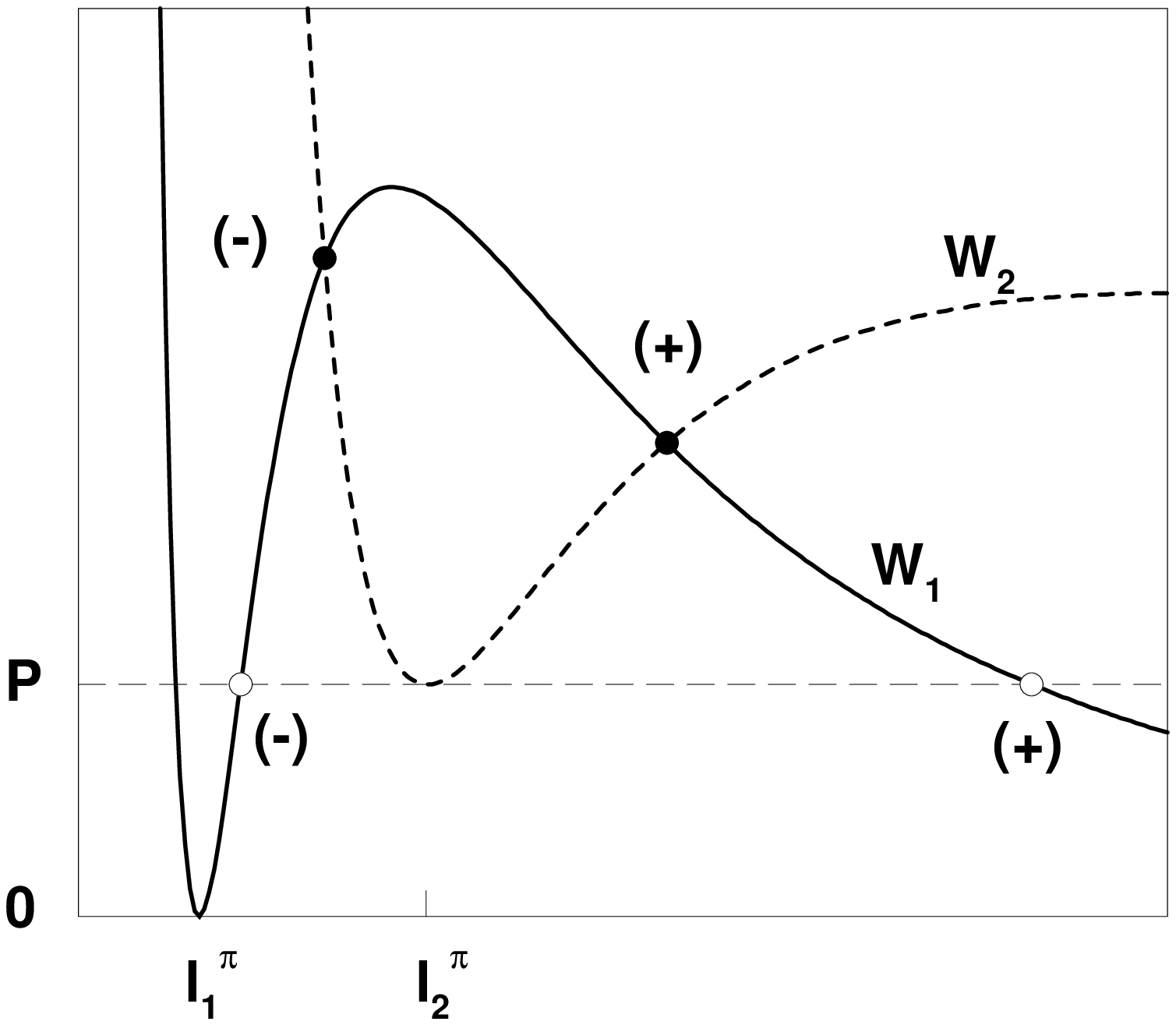,width=7.cm}}
\vspace*{.25cm}  
\caption{Geometrical construction to solve Eq.\ 
(\protect\ref{app3}) for arbitrary effective potentials
$W_1$ and $W_2$ at $T\!=\!T_{\mathrm{w}}^{(1)}$.
$W_1$ shows an activation barrier, necessary to
model a first-order wetting transition.
There are two possible solutions for both $\ell_0$ (empty circles)
and $\ell_L$ (black circles). These correspond to bound (-) and
near-unbound (+) interfacial configurations.}
\end{figure}


\begin{references}
\bibitem{Review} For a review, see, for example, 
S.\ Dietrich, in {\it "Phase Transitions and Critical
Phenomena"}, (C.\ Domb and J.L.\ Lebowitz, eds.), Vol.\ {\bf 12}, p.\ 1
(Academic Press, London, 1988).
\bibitem{Microfluidics} See, for example, B.\ H.\ Weigl, P.\ Yager,
Science {\bf 283}, 346 (1999). See also the collected papers
J.\ Micromech.\ Microeng.\ {\bf 4}, n.\ 4 (1994).
\bibitem{Geometry} C.\ Rasc\'{o}n and A.O.\ Parry,
Nature {\bf 407}, 986 (2000).
\bibitem{Drelich} J.\ Drelich {\it et al.\ }, Langmuir {\bf 12}, 1913
(1996).
\bibitem{Swain} P.S.\ Swain and R.\ Lipowsky, Langmuir {\bf 14}, 6772
(1998).
\bibitem{Wood} J.T.\ Woodward, H.\ Gwin and D.K.\ Schwartz, 
Langmuir {\bf 16}, 2957 (2000).
\bibitem{Adao} M.H.\ Ad\~ao {\it et al.\ }, Phys.\ Rev.\ E {\bf 59},
746 (1999).
\bibitem{Lenz} P.\ Lenz and R.\ Lipowsky, Phys.\ Rev.\ Lett.\ {\bf 80},
1920 (1998).
\bibitem{BD1} C.\ Bauer and S.\ Dietrich,
Eur.\ Phys.\ J.\ B {\bf 10}, 767 (1999).
\bibitem{Urban} D.\ Urban, K.\ Topolski and J.\ de Coninck, 
Phys.\ Rev.\ Lett.\ {\bf 76}, 4388 (1996).
\bibitem{Henderson} J.R.\ Henderson, Mol.\ Phys.\ {\bf 98}, 677 (2000).
\bibitem{PMR} A.O.\ Parry, E.D.\ Macdonald and C.\ Rasc\'{o}n,
J.\ Phys.: Condens. Matter. (to appear), {\tt cond-mat/0008036}.
\bibitem{Palacin} S.\ Palacin {\it et al.\ }, Chem.\ Matter.\ {\bf 8},
1316 (1996).
\bibitem{Frink} L.J.\ Douglas Frink and A.G.\ Salinger, J.\ Chem.\
Phys.\ {\bf 110}, 5969 (1999).
\bibitem{Gau} H.\ Gau {\it et al.\ }, Science {\bf 283}, 46 (1999).
\bibitem{Polymer1} M.\ B\"oltau {\it et al.\ }, Nature {\bf 391}, 877 (1998).
\bibitem{Polymer2} A.\ Karim {\it et al.\ }, Phys.\ Rev.\ E {\bf 57},
R6273 (1998).
\bibitem{Kargupta} K.\ Kargupta, R.\ Konnur and A.\ Sharma,
Langmuir {\bf 16}, 10243 (2000).
\bibitem{BD2} C.\ Bauer and S.\ Dietrich,
Phys.\ Rev.\ E {\bf 60}, 6019 (1999).
\bibitem{BDP} C.\ Bauer, S.\ Dietrich and A.O.\ Parry,
Europhys.\ Lett., {\bf 47} 474 (1999).
\bibitem{BD3} C.\ Bauer and S.\ Dietrich,
Phys.\ Rev.\ E {\bf 61}, 1664 (2000).
\bibitem{Granada} C.\ Rasc\'{o}n and A.O.\ Parry,
J.\ Phys.: Condens. Matter., {\bf 12} A369 (2000).
\bibitem{Weff} Details about effective potentials can be 
found in \protect\cite{Review}.
\bibitem{RPS} C.\ Rasc\'{o}n, A.O.\ Parry and A.\ Sartori,
Phys.\ Rev.\ E {\bf 59}, 5697 (1999).
\bibitem{Flat} The structure of such near-flat profiles
beyond the interfacial model (\protect\ref{two}),
is discussed in ref.\ \protect\cite{BD3}.
\bibitem{Schick} E.H.\ Hauge and M.\ Schick,
Phys.\ Rev.\ B {\bf 27}, 4288 (1983).
\bibitem{First} Standard finite-size scaling theory
predicts that fluctuation effects, beyond the present mean-field
consideration, modify this to a sharply rounded pseudo-first-order
phase transition. For details, see ref.\ \protect\cite{BDP}.
\bibitem{Met} The possibility of using a maximum of the
effective potential to stabilise a solution is discussed
in \protect\cite{Granada}.
\bibitem{Cassie} A.B.D.\ Cassie, Discuss.\ 
Faraday Soc.\ {\bf 3}, 11 (1948).
\bibitem{Israel} J.\ Israelachvili, {\it Intermolecular \& Surface Forces}
(Academic, London, 1991).



\end{references}
\end{document}